\documentclass[11pt]{article}
\usepackage{graphicx}
\usepackage{amsfonts}
\usepackage{amsmath}
\usepackage{amssymb}
\usepackage{url}
\usepackage{fancyhdr}
\usepackage{indentfirst}
\usepackage{enumerate}
\usepackage{dsfont,footmisc}

\allowdisplaybreaks
\usepackage{amsthm}
\usepackage{color}
\usepackage{natbib}
\usepackage[british]{babel}
\usepackage[colorlinks=true,citecolor=blue]{hyperref}
\usepackage[right,pagewise,displaymath, mathlines]{lineno}

\usepackage[font=scriptsize]{caption} 
\usepackage{soul,xcolor}
\setstcolor{red}

\usepackage[title]{appendix}

\def\d{\mathrm{d}}

\newcommand{\VaR}{\mathrm{VaR}}

\newcommand{\ES}{\mathrm{ES}}
\newcommand{\E}{\mathbb{E}}
\newcommand{\R}{\mathbb{R}}

\newcommand{\id}{\mathds{1}}

\newcommand{\X}{\mathcal X}

\renewcommand{\H}{\mathbb H}

\newcommand{\esssup}{\mathrm{ess\mbox{-}sup}}
\newcommand{\essinf}{\mathrm{ess\mbox{-}inf}}
\renewcommand{\ge}{\geqslant}
\renewcommand{\le}{\leqslant}
\renewcommand{\geq}{\geqslant}
\renewcommand{\leq}{\leqslant}
\renewcommand{\epsilon}{\varepsilon}

\theoremstyle{plain}
\newtheorem{theorem}{Theorem}
\newtheorem{corollary}{Corollary}
\newtheorem{lemma}{Lemma}
\newtheorem{proposition}{Proposition}

\theoremstyle{definition}
\newtheorem{definition}{Definition}
\newtheorem{example}{Example}

\theoremstyle{remark}

\theoremstyle{definition}

\renewcommand{\cite}{\citet}

\usepackage[onehalfspacing]{setspace}

\usepackage{geometry}

\begin{document}
\title{Optimal insurance with mean-deviation measures}

\author{Tim J. Boonen\thanks{Department of Statistics and Actuarial Science, The University of Hong Kong, Hong Kong, P.R.China. E-mail: \texttt{tjboonen@hku.hk}} \and
Xia Han\thanks{Corresponding author: School of Mathematical Sciences and LPMC,  Nankai University, Tianjin, P.R.China.   E-mail: \texttt{xiahan@nankai.edu.cn}}
 }


\maketitle
\begin{abstract}
This paper studies an optimal insurance contracting problem in which the preferences of the decision maker given by the sum of the expected loss and a convex, increasing function of a deviation measure. As for the deviation measure, our focus is on convex signed Choquet integrals (such as the Gini coefficient and a convex distortion risk measure minus the expected value) and on the standard deviation. We find that if the expected value  premium principle is used, then stop-loss indemnities are optimal, and we provide a precise characterization of the corresponding deductible. Moreover, if the premium principle is based on Value-at-Risk or Expected Shortfall, then a particular layer-type indemnity is optimal, in which there is coverage for small losses up to a limit, and additionally for losses beyond another deductible. The structure of these optimal indemnities remains unchanged if there is a limit on the insurance premium budget. If the unconstrained solution is not feasible, then the deductible is increased to make the budget constraint binding. We provide several examples of these results based on the Gini coefficient and the standard deviation.
    \end{abstract}
\section{Introduction}
Optimal insurance contract theory has gained substantial academic interest in recent years. In this theory, a decision maker (DM) or policyholder optimizes an objective function based on his/her terminal wealth, and insurance is priced using a well-defined premium principle. Early contributions to this problem studied expected utility \citep{A63} or a mean-variance function \citep{B60} as objectives for the DM. More recent papers study more sophisticated objectives based on regulations or decision-theoretic frameworks that have gained popularity in behavioral economics. To list a few examples, researchers have considered distortion risk measures \citep{CYW13,A15}, expectiles \citep{CW16},  rank-dependent utilities \citep{G19,XZZ19,LWV22}, regret-based objectives \citep{CZ22} and objectives with narrow framing \citep{Z20,CZZ22,LJZ23}. In this paper, our focus is on an objective that is new in the context of optimal insurance contract theory: mean-deviation measures. 

The class of generalized deviation measures was introduced by \cite{RUZ06} via a set of four axioms. It is characterized based on a modified set of axioms compared to \cite{ADEH99}; in particular the translation invariant property of a risk measure $\rho$ is modified from $\rho(X+c)=\rho(X)+c$ \citep{ADEH99}, which is also called cash additivity,  to the translation invariance property: $\rho(X+c)=\rho(X)$, for all random variables $X$ and $c\in\mathbb{R}$. This allows for a natural separation between the actuarial value of a loss (expected loss) and the risk of a loss (measured via deviation measures). Canonical examples of deviation measures include the Gini coefficient and the standard deviation. The Gini coefficient ranges from 0 to 1, where 0 represents no risk and 1 represents the case in which all losses are concentrated in one state of the world (maximum dispersion in a distribution). It measures the extent to which the distribution deviates from a constant, and can be used to measure the risk of a random variable. The standard deviation is a more classical way to measure risk, and its use as risk measure is very popular for Gaussian distributions. 

We study a special class of preferences, which can be seen as a generalization of mean-variance optimization in \cite{M52}. In mean-variance optimization, an individual seeks to find a balance between expected return (mean) and risk (measured using the variance). A key advantage of mean-variance optimization is its simplicity, and this mean-variance structure allows us to explicitly reflect an individual's tolerance towards  risk. In this paper, we keep such simple trade-off structure, and replace the variance is function to measure risk by a more general deviation measure. Deviation measures preserve two key properties of variance: they are non-negative and translation invariant. In this way, deviation measures non-negative risk, and a deterministic loss is measured as zero risk. Deviation measures are consequently used to measure the ``risk'' within a mean-risk trade-off. While variance is not a special case of a deviation measure as it is not positively homogeneous and sub-additive, the standard deviation is. The objective that we study is called mean-deviation measures, and it considers the sum of a convex function of a deviation measure plus the expectation. By taking the square-function as convex function, the original mean-variance objective as in \cite{M52} is recovered as special case. Some specific mean-deviation  measures have been extensively  studied   in the literature on portfolio selection problems,   where the objective is  to minimize the  risk of a  portfolio  subject to a desired expected return,  or   to maximize the return among all portfolios with the risk not exceeding some threshold; see for example   \cite{S64}, \cite{RUZ07} and \cite{RU13}.  Mean-deviation  measures are also studied in the context of risk measures, see for example the   mean-semideviation   in  \cite {OR01}, mean-distortion risk measure mixtures in \cite{DP08} and \cite{CL17} and  mean-Expected Shortfall mixtures in \cite{EMWW21}. For a general study on properties of mean-deviation risk measures, we refer to \cite{HWW23}.

If the premium principle is based on the well-known expected value premium principle, then we find that stop-loss indemnities are optimal if the deviation measure used is a convex signed Choquet integral or the standard deviation. This finding still holds true even if the insurance premium is constrained by a constant budget. This result provides further evidence of the desirability of stop-loss insurance indemnities. This is well-known in the context of mean-variance optimization \citep{B60} and expected utility \citep{A63}, but we show that this holds true in a class of mean-deviation measures. In practice, stop-loss insurance is for insurance provided in public health insurance in the Netherlands, where participants need to pay their healthcare costs in a year up to a given deductible. If the premium principle is based on Value-at-Risk or Expected Shortfall, we show in this paper that the optimal indemnity is generally a dual truncated stop-loss indemnity. In such indemnity function, there is coverage for small losses up to a limit, and additionally for losses beyond another deductible. 



This paper is structured as follows. In Section \ref{sec:2} we formulate the precise problem that we study in this paper. Section \ref{sec:EVPP} presents our main results with the expected value premium principle. Section \ref{sec:DPP} examines two special cases with a distortion premium principle. Section \ref{sec:5} examines the impact of a premium budget constraint, and 
Section \ref{sec:conclusion} concludes. Appendix \ref{sec:AppA} provides some background axioms of risk measures that are referred to in this paper, and Appendix \ref{sec:AppB} provides insights into the monotonicity property of mean-deviation risk measures. Furthermore, Appendix \ref{sec:AppC} provides a proof that was omitted from the main text.
\section{Problem formulation}\label{sec:2}

Let $(\Omega, \mathcal{F}, \mathbb{P})$ be an atomless probability space, and $\mathcal{L}^p, p \in[1, \infty)$ be the set of all random variables with finite $p$-th moment and $\mathcal{L}^{\infty}$ be the set of essentially bounded random variables.   Each random variable represents a random risk that is realized at a well-defined future period. 
 Throughout the paper,  ``increasing" and ``decreasing" are in the  nonstrict (weak) sense, and all functionals we encounter  are law-invariant (see Appendix \ref{sec:AppA} for the definition). Let  $\X$ be a convex cone of random variables.   For any $Z \in \X$, the cumulative distribution function associated with $Z$ is denoted by $F_Z$. For any subset of $I$, we define $\inf{\emptyset}=\esssup\{x: x\in I\}$ and $\sup{\emptyset}=\essinf\{x: x\in I\}$.

 In decision-making,  deviation measures 
 are  introduced and studied  systematically for their application to risk management in areas like portfolio optimization and engineering.  Roughly speaking, deviation measures evaluate
the degree of nonconstancy in a random variable, i.e., the extent to which outcomes may deviate from
expectations. One example of such measures is the standard deviation (SD), which can be considered as a special case. Deviation measures need not be symmetric with respect to ups and downs.     Fix $p\in[1,\infty]$. A mapping  $D: L^p \rightarrow \mathbb{R}$,    is called  \emph{generalized deviation measures} \citep[see, e.g.,][]{RUZ06} if it satisfies 
\begin{itemize}
\item[(D1)] (Translation invariance) $D (Z+c)= D (Z)$ for all $Z\in L^p$ and $c\in\R$.
\item[(D2)] (Nonnegativity) $D (Z)\geq0$ for all $Z\in L^p$, with $ D (Z)>0$ for nonconstant $Z\in L^p$.
\item[(D3)] (Positive homogeneity) $D (\lambda Z)=\lambda  D (Z)$ for all $Z\in L^p$ and all $\lambda\geq0$.
\item[(D4)] (Sub-additivity) $D (Y+Z) \leq  D (Y)+ D (Z)$ for all $Y,Z\in L^p$.
\end{itemize}

 We can see that  the combination of   (D3) with (D4) implies convexity,  thus   $D$ is a convex functional (see Appendix \ref{sec:AppA} for the definition). The set of generalized deviation measures includes, for instance, SD,   semideviation, Expected Shortfall (ES)  deviation and range-based deviation; see Examples 1 and 2 of  \cite{RUZ06} and Section 4.1 of \cite{GMZ12}.  Note that variance does not belong to the generalized deviation measures since it is not positive homogeneous.  For more discussions and interpretations of these properties, we refer to \cite{RUZ06}.  The  continuity of $D$
on $\mathcal L^p$ is defined respect  to $\mathcal{L}^p$-norm.  We denote  $\mathcal{D}^p$ as the set of continuous generalized deviation measures.

Deviation measures are not  risk measures in the sense of  \cite{ADEH99}, but the connection between deviation measures and risk measures is strong. It is shown in Theorem 2 of  \cite{RUZ06} that under some bounded conditions, the generalized deviation measures  correspond one-to-one with coherent risk measures $\rho$ with the relations  that  $D(Z)=\rho(Z)-\E[Z]$ or $\rho(Z)=D(Z)+\E[Z]$ for any $Z\in\X$. Note that the additive structure $\rho=D+\E$ is only as  a special form of the combination of mean and deviation.

In the following definition, we state the  mean-deviation (MD) preferences that we study in this paper.
 \begin{definition}
\label{def:1}Fix $p\in[1,\infty]$, and  let  $D\in\mathcal{D}^p$. A mapping $\mathrm{MD}^D_g : L^p  \to \R$  is defined by   \begin{equation}\label{eq:MD}\mathrm{MD}^D_g(Z)=g(D(Z))+\E[Z],\end{equation}
where   $g:[0, \infty) \rightarrow \mathbb{R}$  is  some continuous, strictly   increasing and convex function  with $g(0)=0$. We use $ \mathcal{G}$ to denote the set of functions $g$.
\end{definition} In this paper, we  aim to study the optimal insurance problems under $\mathrm{MD}^D_g$ on  $\X=\mathcal L^p$ for some  fixed $p\in[1,\infty]$ such that $\mathrm{MD}^D_g$ is finite. 
Note that   $\mathrm{MD}^D_g$ is a convex risk measure since  the
 expectation is linear and $D$ is  convex. Also, since $g$ is strictly  increasing, $\mathrm{MD}^D_g$ yields an aversion towards the deviation of $Z$, as measured by $D(Z)$. The monotonicity of   $\mathrm{MD}^D_g$ is discussed in Appendix \ref{sec:AppB}.

Now, we consider an application of the representation \eqref{eq:MD} in optimal insurance design problems. Suppose that a DM faces a random loss $X\in\X_+$, where $\X_+=\{\X\in\X, X\geq0\}$.  The survival function  $S_X(x)$  of $X$  is assumed to be continuous and strictly decreasing on $[0,M]$, where 
 $M$ is the essential supremum of $X$ (may be finite or infinite). Under an insurance contract, the insurer agrees to cover a part of the loss $X$ and requires a premium in return. The function $I: [0,M]\setminus\{\infty\}\to [0,M]\setminus\{\infty\}$   is commonly described as the indemnity or ceded loss function, while $R(x) \triangleq x-I(x)$ is known as the retained loss function.  
To prevent the potential ex post moral hazard, where the DM might be incentivized to manipulate the size of the loss,  and impose the incentive compatibility condition on the indemnity functions.  We consider insurance contract $I \in \mathcal{I}$, where
\begin{equation}\label{eq:I}
\mathcal{I}:=\left\{I: [0,M]\setminus\{\infty\} \rightarrow [0,M]\setminus\{\infty\} \mid I(0)=0 \text { and } 0 \leqslant I(x)-I(y) \leqslant x-y, \text { for all } 0 \leqslant y \leqslant x\right\} . 
\end{equation}
Obviously, for any $I \in \mathcal{I}, I(x)$ and $x-I(x)$ are increasing in $x$. The assumption that $I\in\mathcal I$ is common in the literature; see, e.g., \cite{A15} and the review paper by \cite{CC20}.

Any $I\in\mathcal I$ is 1-Lipschitz continuous. Given that a Lipschitz-continuous function is absolutely continuous, it is almost everywhere differentiable and its derivative is essentially bounded by its Lipschitz constant. Therefore, function $I$ can be written as the integral of its derivative, and $\mathcal I$ can be represented as
\begin{equation}\label{eq:marginal}
\mathcal I=\left\{I: [0,M]\setminus\{\infty\} \rightarrow [0,M]\setminus\{\infty\} \mid I(x)=\int_0^x q(t) \d t, ~0 \leq q \leq 1\right\}.
\end{equation}
We introduce the space of marginal indemnification functions as
$$
\mathcal Q=\left\{q: [0,M]\setminus\{\infty\} \rightarrow \mathbb{R}_{+}  \mid 0 \leq q \leq 1\right\}.
$$
 For any indemnification function $I \in \mathcal I$, the associated marginal indemnification is a function $q \in \mathcal Q$ such that
$
I(x)=\int_0^x q(t) d t,~ x \geq 0.
$

For a given $I \in \mathcal{I}$, the insurer prices indemnity functions using $\Pi(I(X))$,  then 
 the risk exposure of the DM after purchasing insurance is given by
 $$T_I(X)= X-I(X)+\Pi(I(X)).$$ 
We assume that the DM would like to use $\mathrm{MD}^D_g$ to measure the risk  and aim to solve the following problem \begin{equation}\label{eq:prob}\min _{I \in \mathcal{I}}\mathrm {MD}_g^D(T_I).\end{equation}
If $\Pi$ is $\|\cdot\|_{p}$-continuous,  the  problem  \eqref{eq:prob} admits an optimal solution $I^* \in \mathcal{I}$. To be more precise, take a sequence $\left\{I_n\right\}_{n=1}^{\infty} \subset \mathcal{I}$ such that 
$$
\lim _{n \rightarrow \infty}g(D(X-I_n(X)))+ \E[X]+\Pi[I_n(X)]=\inf _{I \in \mathcal{I}}\{g(D(X-I(X)))+ \Pi(I(X))\}.
$$
Since there exists a subsequence $\left\{I_{n_k}\right\}_{k=1}^{\infty}$ that uniformly converges to $I^* \in \mathcal{I}$,  we know  $I_{n_k}(X) \rightarrow$ $I^*(X)$ in $L^p$ as $k \rightarrow \infty$. Since $D$ and $\Pi$ are $\|\cdot\|_{p}$-continuous, and $g$ is continuous,  then $I^*$ is a minimizer for \eqref{eq:prob}.
Note that  continuity is a technical condition commonly satisfied by most risk measures. For instance, VaR is continuous on $L^{\infty}$ whereas ES is continuous on $L^1$. Below, we consider a set $\X$ such that both $D$ and $\Pi$ are continuous.

\section{Results under expected value premium principle}\label{sec:EVPP}
In this section, we assume that the insurer prices indemnity functions using a premium principle defined as the expected value premium principle:   \begin{equation}\label{eq:ev}\Pi(I(X))=(1+\theta)\E[I(X)],\end{equation} where $\theta>0$ is the safety loading parameter. 

\subsection{Optimal solutions with convex signed Choquet integrals}\label{sec:3.1}
To find an explicit solution  of \eqref{eq:prob},  we  focus on a subset of generalized deviation measures $\mathcal D^p$  by assuming that $D$ is  a  convex signed Choquet integral.  Denote by  $$\widetilde{\mathcal{H}}_c=\{h: h ~\text{is~a ~mapping from}~~ [0,1] ~\text{to}~ \mathbb{R} ~\text{with}~ h(0)=0, ~h(1)=c\}$$ with $c=0$ or $c=1$.  Let 
$$ \rho^c_h(X)=\int_0^\infty  h(S_X(x))\mathrm{d} x, $$ where $h\in\widetilde{\mathcal{H}}_c$.   The function $h$ is called the distortion function of $\rho^c_h$.   For $X\in \X$ with its distribution function given by $F$,  the  Value at Risk (VaR) of $X$ at level $p\in(0,1]$ is  defined as,  for $x \in \mathbb{R}$,
\begin{equation}\label{eq:VaR}\VaR_p(X)=F^-_X(p)=\inf \left\{x\in \R: F(x) \ge p\right\},\end{equation}  which is the left-quantile of $X$. 
It is useful to note that $\rho^c_h$ admits a quantile representation as follows 
 \begin{equation}\label{eq:quantile}\rho^c_h(X)=\int_{0}^{1} \VaR_{1-p}(X) \d  h(p).\end{equation}
   By Theorem 1 of \cite{WWW20}, we know that if $h$ is concave, then $\rho^c_h$ is  convex and comonotonic additive (see Appendix \ref{sec:AppA} for definitions). In the following, we use $\mathcal H_c$ to denote the subset of $\widetilde{\mathcal H}_c$ where $h$ is also concave.  For $h\in \mathcal{H}_c$,  $\rho^c_h$ is finite on $\mathcal{L}^p$ for $p \in[1, \infty]$ if and only if $\|h^{\prime}\|_q<\infty$, where $\|h^{\prime}\|_q=(\int_0^1|h^{\prime}(t)|^q \d t)^{1 / q}$ and $q=(1-1 / p)^{-1}$, and  $\rho^c_h$ is always finite on $\mathcal{L}^{\infty}$; see Lemma 2.1 of \cite{LCLW20}.
   
  There has been an extensive literature on a subclass of signed Choquet integrals, in
which $h\in\widetilde{\mathcal H}_1$ is increasing; we  call this class of functionals  distortion risk measures.  Further, the signed Choquet integrals are also  used as measures of
distributional variability, where  $h\in \widetilde{\mathcal H}_0$. In this case,  $h$ is  not monotone.  Note that $\rho_h^0$  with $h\in  \mathcal H_0$ satisfies all the four properties of (D1)-(D4), and thus belong to the class of the  generalized deviation measures. 
 In particular,  by Theorem 1 of \cite{WWW20}, if  a generalized deviation measure $D$ is comonotonic additive,   then $D$  can only be the signed Choquet integrals.  When  $h\in \mathcal H_0$, we refer to Appendix \ref{sec:AppA} for more specific examples.

 Thus, when $D$ is a  signed Choquet integral,  our objective in \eqref{eq:prob} can be rewritten as 
\begin{equation}\label{eq:prob11}\min _{I \in \mathcal{I}}\mathrm {MD}_g^D(T_I)= \min _{I \in \mathcal{I}}\left\{ g(D_h(X)-D_h(I(X)))+ \E[X]+\theta\E[I(X)]\right\},\end{equation}  where  \begin{equation}\label{eq:D_h}
      D_h(X):= \rho^0_h(X)=\int_0^\infty  h(S_X(x))\mathrm{d} x, \end{equation} with $h\in \mathcal H_0$.\footnote{ We remark that all convex signed Choquet integral on $\mathcal{L}^p$ are $\mathcal{L}^p$-continuous; see Corollary 7.10 in \cite{R13} for the $\mathcal{L}^p$-continuity of the finite-valued convex risk measures on $\mathcal{L}^p$.} This is a direct consequence of $D_h(X-I(X))=D_h(X)-D_h(I(X))$, which is due to comonotonic additivity of $D_h$.  

\medskip

\begin{theorem}\label{thm:1}Suppose that $D$ is given by \eqref{eq:D_h} and $\Pi$ is given by \eqref{eq:ev}. The following statements hold: 
\begin{enumerate}[(i)]
    \item  For every $I\in\mathcal I$,
 we can construct a stop-loss insurance treaty $  I_d(x)=(x-d)_+$ for some $0\leq d\leq M$ such that $\mathrm{MD}^D_g (T_{I_d})\leq \mathrm{MD}^D_g (T_{I})$. Further, $  I_{d^*}$ with  \begin{equation}\label{eq:d1} d^*=\sup\left\{x:~g'\left(  \int _0^{ x } h(S_X(t))  \d t\right) h(S_X(x))-\theta S_X(x)\leq 0,~\text{and}~0\leq x< M\right\},\end{equation} is a solution to problem \eqref{eq:prob}.
 \item If  $h''(0)<0$, the optimal solution  to problem \eqref{eq:prob}   is unique on  $[0,M]$,  i.e., we have $I_d^*=\arg\min _{I \in \mathcal{I}} \mathrm{MD}^D_g (T_{I})$.
\end{enumerate}
  \end{theorem}
\begin{proof}  To show (i), we first fix $D_h(X-I(X))=s\in [0,D_h(X)]$ and solve \eqref{eq:prob} subject to this constraint. That is, we want to solve 
\begin{equation}\label{eq:f1}  \min _{I \in \mathcal{I}} f(I):=g(s)+\theta \E[I(X)]+ \E[X]+\lambda (D_h(X-I(X)))-s),\end{equation}  where   $\lambda\geq0$ being the Karush-Kuhn-Tucker (KKT) multiplier. By   \eqref{eq:marginal} and \eqref{eq:quantile},  we have  
\begin{align*}f(I)&=\theta \int _0^1  \VaR_{1-t}(I(X))\d t+\lambda \int _0^1 \VaR_{1-t}(X-I(X)) \d h(t) + g(s)+ \E[X]-\lambda s
\\ &=\theta \int _0^1  I(\VaR_{1-t}(X))\d t+\lambda \int _0^1(  \VaR_{1-t}(X) -I(\VaR_{1-t}(X)))  \d h(t) + g(s)+ \E[X]-\lambda s\\
&=\theta \int _0^M S_X(x) q(x) \d x +\lambda \int _0^M h(S_X(x)) (1-q(x)) \d x + g(s)+ \E[X]-\lambda s\\& =  \int _0^M( \theta S_X(x)-\lambda h(S_X(x)))  q(x) \d x   + \lambda D_h(X)+ g(s)+  \E[X]-\lambda s.\end{align*}  
The second equality follows from comonotonic additivity of $\VaR$ and  $f(\VaR_t(X))=\VaR_{t}(f(X))$ for any increasing function $f$ and $t\in(0,1)$, and the third equality follows from a change of variable and integration by parts.  Define 

$$\underline d_{\lambda} =\sup\{x:~\theta S_X(x)-\lambda h(S_X(x))> 0~\text{and}~0\leq x< M\},$$ 
and  
$$\overline d_{\lambda} =\sup\{x:~\theta S_X(x)-\lambda h(S_X(x))\geq0~\text{and}~0\leq x< M\}.$$ 
 It is obvious that $\underline d_{\lambda}\leq\overline d_{\lambda} $  for any fixed $\lambda\in[0,\infty)$. Define  $H(x)= \theta S_X(x)-\lambda h(S_X(x))$. 
 It  is clear that  $H(0)=\theta>0$, $\lim_{x\to M}H(x)=0$, and   $H'(x)= (\theta -\lambda h'(S_X(x)))S'_X(x).$  Since $h$ is a concave function with $h(0)=h(1)=0$, if $\lambda <\theta/h'(0)$ (i.e., $H'(M)<0$),  we have $q(x)=0$ and $ \underline d_\lambda=\overline d_\lambda =M$, and thus $I(x)=0$. Otherwise, if $\lambda \geq \theta/h'(0)$, 
 it is clear that the following $q$ will minimize \eqref{eq:f1}
\begin{equation}\label{eq:q1}q(x)=\left\{\begin{aligned}&0,~~~~~&\text{if }\theta S_X(x)-\lambda h(S_X(x))>0~~ (i.e.,  ~x< \underline d_\lambda ),\\&1,~~~~~&\text{if }\theta S_X(x)-\lambda h(S_X(x))<0~~ (i.e.,  ~x> \overline d_\lambda ),\\& c(x),~~~&\text{otherwise},\end{aligned}\right.\end{equation}
where $c(x)$ could be any $[0,1]$-valued function.  
Thus, we can select the function $c$ to be of the form $c(x)=1_{\{x>d\}}$ for some $d\in[\underline d_\lambda,\overline d_\lambda]$. Then, $I(x)=I_d(X):=\int_0^x q(t)\d t =(x-d)_+$. Now, $\lambda$ is such that $D_h(X-I_{\underline d_\lambda}(X))\geq s$ and $D_h(X-I_{\overline d_\lambda}(X))\leq s$, and since $D_h(X-I_{d}(X))$ is increasing in $d$,  there exists $d\in[\underline d_\lambda,\overline d_\lambda]$ such that 
$$s=D_h(X-I_d(X))=  \int _0^{d} h(S_X(x))  \d x.$$ That is,  for every $s$,  there exists an $I_d(x)=(x-d)_+$ that does better than any $I\in\mathcal{I}$. 

We next show that for any $I^*$ that solves  
\eqref{eq:prob},  there exists an $I_d(x)=(x-d)_+$ such that $\mathrm{MD}^D_g(T_{I^*})=\mathrm{MD}^D_g(T_{I_d})$.  We fix $I^*$ that solves  
\eqref{eq:prob}, and define $s=D_h(X-I^*(X))$. By the above steps, for any given $s$,   we can always construct an insurance treaty $ {I}_d(x)=(x-d)_+$ for some $0 \leq d \leq M$ such that $ \mathrm{MD}^D_g(T_{I_d})\leq \mathrm{MD}^D_g(T_{I^*})$.      Since $I^*$  is optimal, then we have $\mathrm{MD}^D_g(T_{I^*})=\mathrm{MD}^D_g(T_{I_d})$. Hence, there exists an optimal indemnity that is of a stop-loss form. 

Finally, we  aim to find the optimal $d$ for  problem  \eqref{eq:prob} by assuming that the insurance contract is given by $I_d$ for some $d\in[0,M]$, that is, 
\begin{equation}\label{eq:F1}\min _{d \in [0,M]} F(d):=g\left( \int _0^{d} h(S_X(x))  \d x\right)+\theta \int _{d}^M S_X(x)  \d x + \E[X]. \end{equation}
 To find the optimal $d$,  with the first-order condition, we have $$F'(d)=g'\left( \int _0^{d} h(S_X(x))  \d x\right) h(S_X(d))-\theta S_X(d) .$$
It is clear that $F'(0)=-\theta<0$ and $F'(M)=0$. Moreover,  
$$F''(d)=g''\left(  \int _0^{d } h(S_X(x))  \d x\right) h^2(S_X( d ))+ g'\left(  \int _0^{d } h(S_X(x))  \d x\right) h'(S_X( d )) S'_X(d)  -\theta S'_X(d).$$  Since  $g$ is convex, $S_X(d)$ decreases in $d$ and $h$ is concave,  $F''$ has at most one intersection with the
x-axis. Let  $$d^*=\sup\left\{x:~g'\left(  \int _0^{ x } h(S_X(t))  \d t\right) h(S_X(x))-\theta S_X(x)\leq 0,~\text{and}~0\leq x< M\right\},$$ 
then  $d^*$  is the optimal solution to \eqref{eq:F1}. This concludes the proof of (i).

To show (ii), if  $h''(0)<0$, then it holds for any concave function with $h(0)=0$ that $h(s)/s$ is strictly decreasing, and since $S_X$ is strictly decreasing on $[0,M]$, therefore it holds that the set $\{x\in[0,M]|~\theta S_X(x)-\lambda h(S_X(x))=0\}$ has Lebesgue measure zero.  In other words, if  $h''(0+)<0$, then  $\underline d_\lambda= \overline d_\lambda$. Then, the necessary condition for optimality of reinsurance contract given by the expression \eqref{eq:d1} becomes a sufficient condition. It implies that $d^*$ is a saddle point  of the function $f(d)$ on $[0,M)$ or $d^*=M$, i.e. $I_{d^*}=\arg\min _{I \in \mathcal{I}} f(I)$.
\end{proof}

In the following corollary,    take $g(t)=\alpha t+\beta t^2$. Then we have  $g'(x)=a+2\beta x$ and $g''(x)=2\beta$. Since $g\in\mathcal G$,  we  assume that $\alpha\geq0$ and  $\beta\geq0$, and at least one of the inequalities holds strictly. 
\begin{corollary}\label{cor:2}
Suppose that $D$ is given by \eqref{eq:D_h} with $h''(0)<0$.
Let  $g(x)=\alpha x+\beta x^2$ with  $\alpha\geq0$ and $\beta\geq0$. 
Then we have  $I^*(x)=(x-d^*)_+,$
where   
 $$ d^* =\sup\left\{x: h(S_X(x))\left(\alpha +2\beta \int _0^{ x} h(S_X(t))  \d t\right)-\theta S_X(x) \leq 0~\text{and}~0\leq x< M  \right\}.$$ 
 In particular, if $\beta=0$, we have $$d^*=\sup\{x:\alpha h(S_X(x))-\theta S_X(x)\leq 0~\text{and}~0\leq x< M\}. $$ 
\end{corollary} 
We remark that $d^*$ in Corollary \ref{cor:2} decreases as $\alpha$ and $\beta$ increase, but increases as $\theta$ increases.   In fact, larger values of $\alpha$ and $\beta$ mean that the DM is more concerned with the variability of the  risk exposure.  Thus, it is to be expected that the DM is willing to buy more insurance when more weight is given to the  deviation.  Specifically,  we have $d^*\to M$ as $\alpha\to0$ and $\beta\to0$, which implies the DM would like to buy no insurance. In this situation, the DM is  risk neutral since $\mathrm{MD}^D_g=\E$.  
 Here, we observe that the quadratic function $g$ can be understood as the DM considering or penalizing the second-order changes in the deviation. Furthermore, as the value of $\theta$ increases, the insurer sets a relatively higher insurance premium, which consequently leads the DM to reduce the amount of insurance purchased.

In the following, we show one special example  by assuming that  $D_h(X)$ is the Gini deviation.  Let $X \in \X$ and  $X_1$, $X_2$, $X$  are i.i.d.,
$$
D_h(X)=\mathrm{Gini}(X):=\frac{1}{2} \mathbb{E}\left[\left|X_1-X_2\right|\right].
$$
The Gini deviation is a signed Choquet integral with a concave distortion function $h$ given by $h(t)=t-t^2, t \in[0,1]$. This is due to its alternative form \citep[see, e.g.,][]{D90}
$$
\mathrm{Gini}(X)=\int_0^1 F_X^{-1}(t)(2 t-1) \mathrm{d} t.
$$ Since $h''(0)=-1<0$ for Gini, by Theorem \ref{thm:1} (ii), $I_d^*$ is   the unique optimal solution. 
\begin{example}\label{exm:1} Let $h(t)=t-t^2$ with $t\in[0,1]$ and $g(x)=\alpha x+ \beta x^2$ with    $\alpha\geq0$ and $\beta\geq0$. If $\beta=0$, then 
$$\theta S_X(x)-\alpha (S_X(x)-S^2_X(x))= S_X(x)(\theta-\alpha +\alpha S_X(x)).$$ Thus, we can see that if $\theta<\alpha$, then $d^*=S_X^-(\frac{\alpha-\theta}{\alpha})$; otherwise, $d^*=M$. 

 For the case of $\beta\neq 0$, we have \begin{align*}&\theta S_X(x) -h(S_X(x))\left(\alpha +2\beta \int _0^{ x} h(S_X(t))  \d t\right)\\&=\theta S_X(x) -(S_X(x)-S^2_X(x))\left(\alpha +2\beta \int _0^{ x} S_X(t)-S^2_X(t)  \d t\right)\\&=S_X(x)\left((\theta-\alpha+\alpha S_X(x)) -2\beta (1-S_X(x)) \int _0^{ x} (S_X(t)-S^2_X(t))  \d t\right). \end{align*} 
If  $X\sim U[a,b]$,\footnote{When $a>0$, the uniform distribution is not covered by Theorem \ref{thm:1} because  $\essinf X$ can be larger than $0$; however, we can modify the proof of Theorem  \ref{thm:1} to account for $X$ with any bounded and non-negative support.} then $\mathrm{Gini}(X)=(b-a)/6$. Take $\theta=0.2$, we can compute $d^*$   numerically by
$$d^*=\sup\left\{x: \frac{b-x}{(b-a)}\left(\alpha\frac{x-a}{b-a}-2\beta  \frac{x-a}{(b-a)^3}\left(\frac{x^3-a^3}{3}-\frac{(a+b)(x^2-a^2)}{2} +a b (x-a)\right)-\theta\right)\leq 0\right\}.$$

\begin{figure}[htb!]
\centering
 \includegraphics[width=15cm]{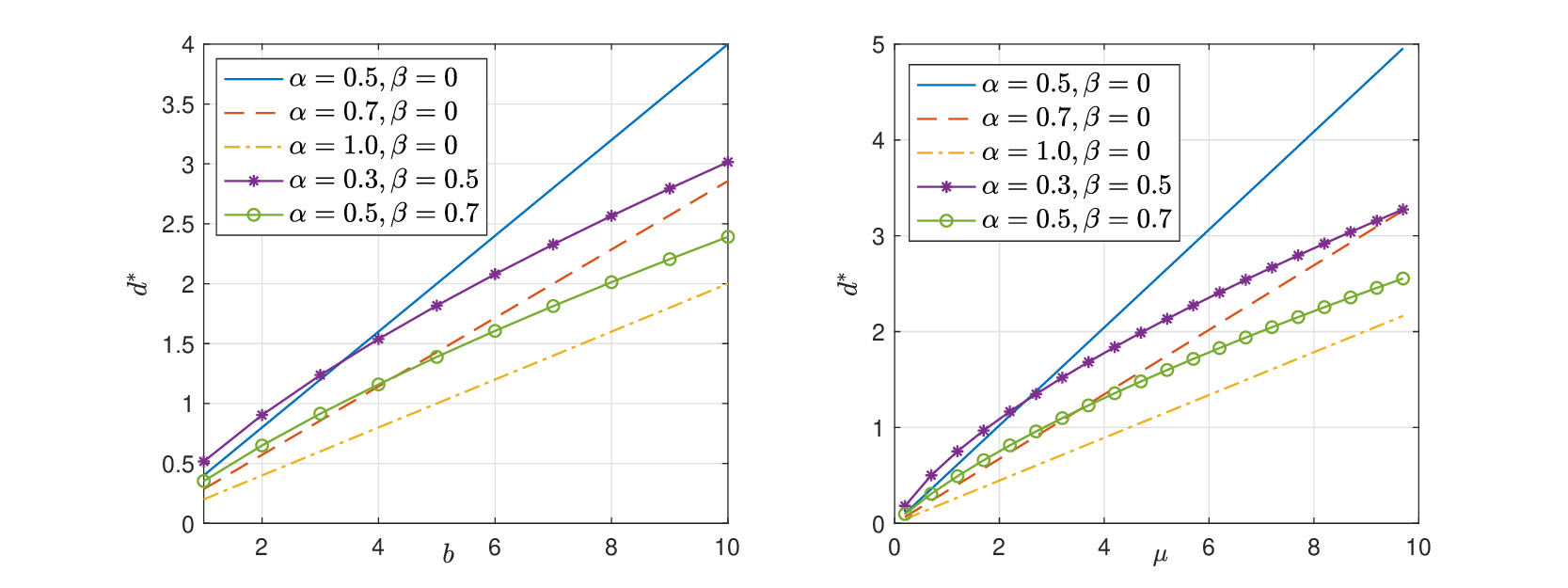} \captionsetup{font=small}
 \caption{ \small Optimal deductible $d^*$ as a function of $b$ for the uniform distribution with $a=0$ (left figure) and as a function of $\mu:=1/\lambda$ for the  exponential distribution (right figure).}\label{fig:1}
\end{figure}

If  $X\sim\exp(\lambda)$ with any $\lambda>0$, then $\mathrm{Gini}(X)=1/(2\lambda )$. Again, we can compute $d^*$   numerically  $$d^*=\sup\left\{x: e^{-\lambda x} (\alpha-\alpha e^{-\lambda x}+\frac{\beta}{\lambda}(1-e^{-\lambda x})^3-\theta)=0\right\}.$$

In Figure \ref{fig:1}, we display the optimal deductible $d^*$ as a function of $b$ for the uniform distribution and as a function of  $\mu:=1/\lambda$ for the  exponential distribution. We find that increasing the expected loss leads to an strict increase in the deductible. This pattern is linear when the function $g$ is linear ($\beta=0$), and concave when the function $g$ is strictly convex ($\beta>0)$. For the uniform distribution, we note that the expected loss before insurance is $b/2$, and the deductible is consistently smaller than $0.4b$. Thus the deductible is paid in full by the DM with a probability that exceeds 0.6. For the exponential distribution, the deductible generally is not larger than $0.5\mu$. Note that here the deductible is paid in full by the DM with a probability that exceeds $\exp(-0.5)\approx0.6$, which is thus similar as for the uniform distribution.
\end{example}

\subsection{Standard deviation based measures}
 As mentioned in Section \ref{sec:2},   SD is a generalized deviation measure, but variance  does not satisfy (D3). Also, neither SD nor variance are convex signed Choquet integrals, so we cannot use Theorem \ref{thm:1} for SD. In particular,  SD can be written  as
$
\mathrm{SD}(X)=\sup \{\int_0^1 \operatorname{VaR}_t(X) \mathrm{d} h(t): h \in \Phi_0,\left\|h^{\prime}\right\|_2^2 \leq 1\},  X \in \mathcal{L}^\infty;
$ see Example 2.1 of \cite{WWW20} for a simple proof of this representation. 

Since  SD and variance are commonly used deviation measures,  we also want to solve \eqref{eq:prob11} with $D=\mathrm{SD}$: \begin{equation}\label{prob:2} \min _{I \in \mathcal{I}}\left\{ g(\mathrm{SD}(X-I(X)))+\E[X]+\theta \E[I(X)]\right\}.\end{equation}  In particular, if $g(x)=\gamma x^2$ for $\gamma>0$, it is the mean-variance criterion. The following lemma is well-known (see, e.g., Property 3.4.19 in \cite{DDGK05} and Lemma A.2 in \cite{C12}).
\begin{lemma} \label{lem:3}Provided that the random variables $Y_1$ and $Y_2$ have finite expectations, if they satisfy
$$
\mathbb{E}\left[Y_1\right]=\mathbb{E}\left[Y_2\right], \quad F_{Y_1}(t) \leq F_{Y_2}(t),\quad t<t_0,~~S_{Y_1}(t) \leq S_{Y_2}(t),\quad t \geq t_0
$$
for some $t_0 \in \mathbb{R}$, then $Y_1 \leq_{c x} Y_2$, i.e.
$$
\mathbb{E}\left[G\left(Y_1\right)\right] \leq \mathbb{E}\left[G\left(Y_2\right)\right]
$$
for any convex function $G(x)$ provided the expectations exist.
\end{lemma}Denote by  $$w_1(d)=\int_0^d  S_X(x)\d x ,~~ \text{and}~~ w_2(d)=2 \int_0^d  x S_X(x)\d x.$$
 \begin{theorem}\label{thm:2}
For problem \eqref{prob:2},  we can construct a stop-loss insurance treaty $\overline  I(x)=(x-d)_+$ for some $0\leq d\leq M$ such that $\mathrm{MD}^D_g (T_{\overline I})\leq \mathrm{MD}^D_g (T_{I})$  for any admissible ceded loss function $I\in \mathcal{I}$. Further, when $g(x)=\alpha x+\beta x^2$ with $\alpha\geq0$ and $\beta\geq0$,    we have $ I^*(x)=(x- d^*)_+$ with  $$ d^*=\sup \left \{x:  \alpha \sqrt {\frac{(x-w_1(x))^2}{w_2(x)-w^2_1(x)} }+2\beta  (x-w_1(x))- \theta \leq0,~\text{and}~0\leq x<M\right\}.$$ 
 \end{theorem}
 \begin{proof} For any admissible ceded loss function $I\in\mathcal I$, we can construct an insurance treaty $\overline {I}(x)=(x-d)_+$ for some $0 \leq d \leq M$ such that $\E[I(X)]=\E[(X-d)_+]$.  Since $k(d):=\E[(X-d)_+]$ is a decreasing function in $d$ and 
with $k(0)=\E[X]$ and $k(M)=0$, together  with $0\leq \E[I(X)]\leq\E[X]$, the existence of $d$ can be verified.   Further, by taking $t_0=d$ in Lemma \ref{lem:3}, we have  $\E[(X\wedge d)^2]\leq \E[(X-I(X))^2]$. Thus, we have  $\mathrm{SD}(X-\overline I(X))\leq \mathrm{SD}(X- I(X))$, which implies that $\mathrm{MD}^D_g (T_{\overline I})\leq \mathrm{MD}^D_g (T_{I})$. 
Therefore,  we have 
 $$\begin{aligned} g(\mathrm{SD}(X\wedge d))+ \E[X]+\theta \E[(X-d)_+]=g\left( \left (w_2(d)-w^2_1(d)\right)^{1/2} \right)+\E[ X] +\theta  \int_d^M  S_X(x)\d x.\end{aligned}$$
Let     $$  f(d)=g(\sqrt{w(d)})+\E[X]+\theta\int_d^M S_X(x) \d x, $$ where $w(d)= w_2(d)- w_1^2(d).$  It is clear that 
$$w'(d)=2dS_X(d)-2 S_X(d)  \int_0^d  S_X(x)\d x=  2 S_X(d) (d-w_1(d)) \geq 0 .$$
 Then we have 
$$ f'(d)= \frac{1}{2 \sqrt {w(d)}}g' ( \sqrt{w(d)} )w'(d)-\theta S_X(d)= S_X(d)\left(   \frac{g' ( \sqrt{ w(d)}) }{\sqrt {w(d)} } (d-w_1(d)) -\theta \right). $$  For $g(x)=\alpha x+\beta x^2$ with $\alpha\geq0$ and  $\beta\geq0$,  we have  $$    F (d):=\frac{g' ( \sqrt{ w(d)}) }{\sqrt {w(d)} }  (d-w_1(d))-\theta=\alpha \sqrt {\frac{(d-w_1(d))^2}{w_2(d)-w^2_1(d)} }+2\beta  (d-w_1(d)) -\theta.$$ 
Let $\phi(d)= \frac{(d-w_1(d))^2}{w_2(d)-w_1(d)}.$ It then follows that   $$\begin{aligned}
\phi'(d)&=\frac{d-w_1(d)}{(w_2(d)-w^2_1(d))^2} {\left(2F_X(d)(w_2(d)-w^2_1(d))-(d-w_1(d)) (w'_2(d)-2w_1(d)w'_1(d))\right)}\\&=\frac{d-w_1(d)}{(w_2(d)-w^2_1(d))^2} {\left(2F_X(d)(w_2(d)-w^2_1(d))-2S_X(d)(d-w_1(d))^2  \right)} \\&\geq \frac{2(d-w_1(d))}{(w_2(d)-w^2_1(d))^2} \left(F_X(d)w_2(d)-(w_1(d)-S_X(d)d)^2\right)\\&= \frac{2(d-w_1(d))}{(w_2(d)-w^2_1(d))^2} \left(F_X(d)S_X(d)d^2+F_X(d)\int _0^d x^2 \d F_X(x)-\left( \int_0^d x \d F_X(x) \right)^2 \right)\\&= \frac{2(d-w_1(d))}{(w_2(d)-w^2_1(d))^2} \left(S_X(d)F_X(d)d^2-S_X(d )\left( \int_0^d x \d F_X(x) \right)^2+F_X(d)\int _0^d x^2 \d F_X(x) \right. \\& \left.~~~-F_X(d)\left( \int_0^d x \d F_X(x) \right)^2 \right)\geq0.\end{aligned}$$ Together with $\lim_{d\to0}\phi(d)=0$, we have $\phi(d)\geq 0$ for $d\in[0,M]$.    It is not difficult to verify that $F(0)=-\theta$ and  $$ F(M)=\left(\frac{\alpha}{\mathrm{SD}(X)}+2\beta\right) (M-\E[X])-\theta. $$ 
Therefore, if $ F(M)<0$, then $ f$ is a decreasing function in $d$ and thus $ d^*=M$. On the other hand,  if $ F(M)\geq0$, $f$ first decreases and then increases in $d$, and thus $$ d^*=\sup \left \{x:  \alpha \sqrt {\frac{(x-w_1(x))^2}{w_2(x)-w^2_1(x)} }+2\beta  (x-w_1(x)) -\theta \leq 0,~\text{and}~0\leq x< M\right\}.$$ 
 \end{proof} Note that for the quadratic function $g(x)=\alpha x+\beta x^2$ it holds that if $\alpha>0$ and $\beta=0$ then $\mathrm{MD}^D_g$ is mean-SD, and if $\alpha=0$ and $\beta>0$ then $\mathrm{MD}^D_g$ is mean-variance.  
 \begin{example}For $X\sim U[0,b]$,  we have $w_1(x)=(2bx-x^2)/{(2b)}$ and $w_2(x)={x^2(3b-2x)}/{(3b)}$. By setting $\theta=0.2$, we can compute $d^*$ numerically by  $$ d^*=\sup \left \{x:  \alpha \left(\frac{3x}{4b-3x}\right)^{1/2}+\frac{\beta x^2}{b}   -\theta \leq 0,~\text{and}~0\leq x\leq b\right\}.$$ 

 For $X\sim\exp(\lambda)$ with any $\lambda>0$,  we have $w_1(x)=(1-e^{-\lambda x})/{\lambda}$ and $w_2(x)=\frac{2}{\lambda^2}(1-e^{-\lambda x})-\frac{2}{\lambda} x e^{-\lambda x}$. By setting $\theta=0.2$, we can compute $d^*$ numerically by  $$d^*=\sup \left \{x:  \alpha \left(\frac{(\lambda x-1+e^{-\lambda x})^2}{1-e^{-2\lambda x}-2\lambda xe^{-\lambda x}}\right)^{1/2}+2 \beta(x-\frac{1-e^{-\lambda x}}{\lambda} )   -\theta \leq 0,~\text{and}~ x\geq0\right\}.$$ 

In Figure \ref{fig:2}, we display  the optimal deductible $d^*$ as a function of $b$ for the uniform distribution and as a function of  $\mu$ for the  exponential distribution. Similar to Figure \ref{fig:1},
we find that increasing the expected loss leads to an strict increase in the deductible. Again, this graph is linear when the function $g$ is linear ($\beta=0$), and concave when the function $g$ is strictly convex ($\beta>0)$. We do find however find that the size of the deductible is substantially smaller than Figure \ref{fig:1}, which is an indication that SD and variance make the DM more risk averse. 
 
 \begin{figure}[htb!]
\centering
 \includegraphics[width=15cm]{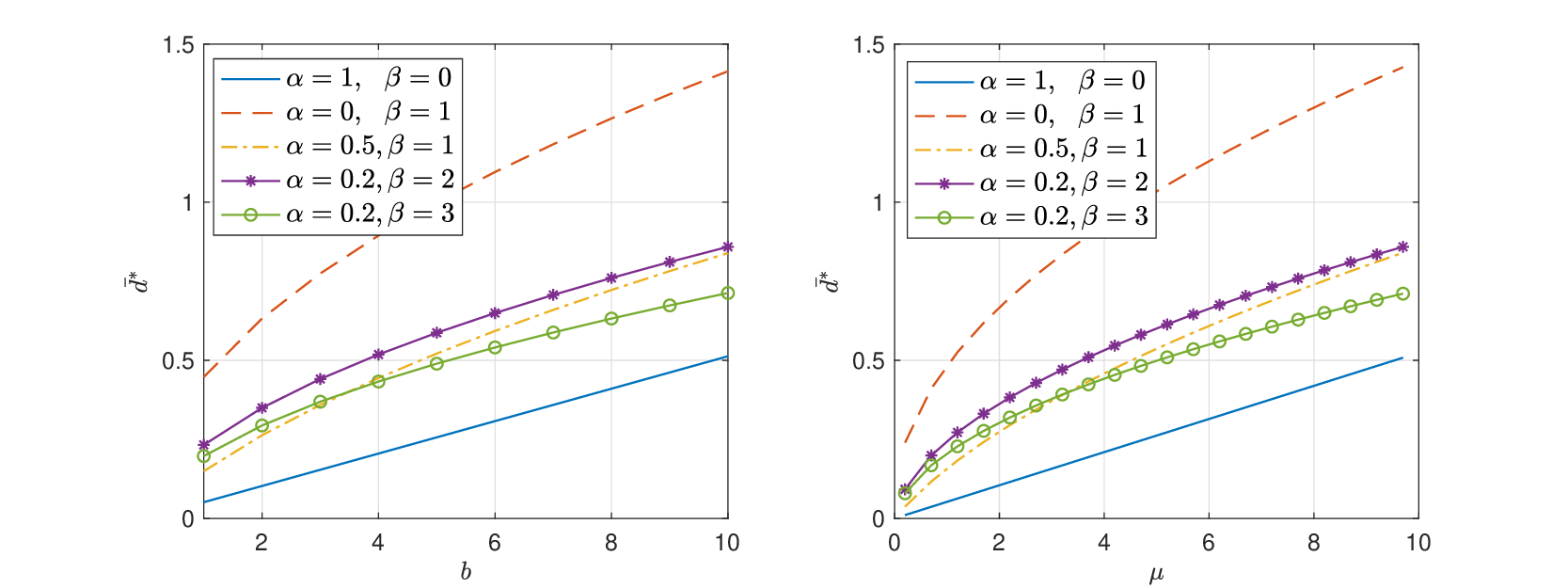} \captionsetup{font=small}
 \caption{ \small Optimal deductible $\overline d^*$ as a function of $b$ for the uniform distribution  (left figure) and as a function of $\mu:=1/\lambda$ for the exponential distribution (right figure).}\label{fig:2}
\end{figure}
 \end{example}

\section{Results for two distortion premium principles}\label{sec:DPP} 

For $h\in\widetilde{\mathcal{H}}_1$ being increasing,  the distortion premium principle $\Pi_{h}$ is given by
\begin{equation}\label{eq:dis_pre}
\Pi_{h}(I(X)):=\int_0^\infty h(S_{I(X)}(x))\d x=\int_0^\infty h(S_{X}(x))q(x)\d x,
\end{equation} where $q$ is defined in \eqref{eq:marginal}, and the second equality above is shown in the proof of Theorem \ref{thm:1}.
When the distortion function $h$ is concave, the amount $\int_0^{\infty}h(S_X(x))  \mathrm{d} t-\mathbb{E}[X]$ is non-negative and can be interpreted as the risk loading that is added to the expected loss. 

In this section,    suppose that $D=D_{h_1}$  with $h_1\in \mathcal H_0$,  we aim to solve 
\begin{equation}\label{eq:prob3}\min _{I \in \mathcal{I}}\mathrm {MD}_g^D(T_I)= \min _{I \in \mathcal{I}}\left\{ g(D_{h_1}(X-I(X)))+ \E[X-I(X)]+\Pi_{h_2}(I(X)) \right\},\end{equation}
where $h_2\in \widetilde{\mathcal H}_1$ is increasing.  As we known, VaR and ES are special distortion risk measures,  where the ES at  level $p\in(0,1)$ is the functional $\ES_p:L^1 \to \mathbb{R}$ defined by
$$
  \ES_p(Z)=\frac{1}{1-p}\int_p^1 \VaR_s(Z)\d s,
$$ where VaR is defined in \eqref{eq:VaR},
 and  $\ES_{1}(Z)=\text { ess-sup }(Z)=\VaR_{1}(Z)$  which may be infinite.   In particular, we have $h(t)=\id_{\{t>1-p\}}$ for $\VaR_p$ and $h(t)=\frac{t}{1-p} \wedge 1$ for $\ES_p$.   The explicit solutions are  derived when the  DM uses VaR  and ES  as the premium principles. For notational convenience, we write $x_p:=\VaR_p(X)$ for some $p\in(0,1)$.

\subsection{Value-at-Risk}
We give the optimal results for $\Pi=\VaR_p$ for $p\in(0,1)$ in the following proposition. 
\begin{proposition}\label{prop:1} Suppose that $D$ is given by \eqref{eq:D_h}, and $h_2(t)=\id_{\{t>1-p\}}$ with $p\in(0,1)$, i.e.,  $\Pi_{h_2}(X)=\VaR_p(X)$. 
The  unique solution to problem \eqref{eq:prob3} is given by $I_{d,x_p}(x)=x\wedge d^*+  (x-x_p)_+$  with  \begin{equation}\label{eq:d_VaR}d^*  =\sup\left\{x:~1- S_X(x)-g'\left(  \int _{x }^{x_p} h_1(S_X(t))  \d t\right) h_1(S_X(x))\leq0,~~\text{and} ~~0\leq x\leq x_p\right\}. \end{equation}
\end{proposition}
\begin{proof} 
 The proof is similar to the one of Theorem \ref{thm:1}, and so we only provide the major steps
that highlight the differences.
We first fix $D_h(X-I(X))=s\in [0,D_h(X)]$ and solve problem \eqref{eq:prob3} subject to this constraint. That is, we want to solve 
$$  \min _{I \in \mathcal{I}} f(I):=g(s)+ \E[X]-  \E[I(X)]+\Pi_{h_2}(I(X))+\lambda (D_{h_1}(X-I(X)))-s)$$ with $\lambda\geq0$ being the KKT multiplier. As shown in  Theorem \ref{thm:1}, $f(I)$ can be written as \begin{align}\label{eq:f5}f(I) =  \int _0^M( h_2(S_X(x))-S_X(x)- \lambda h_1(S_X(x)))  q(x) \d x   + \lambda D_{h_1}(X)+ g(s)+  \E[X]-\lambda s,\end{align}  and  it is clear that the following $q$ will minimize \eqref{eq:f5}:
\begin{equation}\label{eq:general}q(x)=\left\{\begin{aligned}&0,~~~&\text{if }h_2(S_X(x))-S_X(x)- \lambda h_1(S_X(x))>0,\\&1,~~~&\text{if } h_2(S_X(x))-S_X(x)- \lambda h_1(S_X(x))<0 ,\\& c,~~~&otherwise,\end{aligned}\right.\end{equation}where $c$ could be any $[0,1]$-valued function on $h_2(S_X(x))-S_X(x)- \lambda h_1(S_X(x))=0$. 
Define $$H(x)=h_2(S_X(x))-S_X(x)- \lambda h_1(S_X(x)).$$
Let  $h_2(t)=\id_{\{t>1-p\}}$ with $p\in(0,1)$. 
 \begin{itemize}
 \item[(i)] For  $t<1-p$, or equivalently, $x_p<x\leq M$, we always have $H(x)=-S_X(x)- \lambda h_1(S_X(x))\leq0,$ which implies $q(x)=1$ for $x_p<x< M$.  
\item[(ii)]For $t\geq1-p$, or equivalently,  $x\leq x_p$, we have $H(x)=1-S_X(x)-\lambda h_1(S_X(x)).$  Note that $h_1$ is concave with $h_1(0)=h_1(1)=0$, $H'(x)=-S'_X(x)(1+\lambda h'_1(S_X(x))$ and $H(x_p)=p-\lambda h(1-p)$.  When $H(x_p)>0$,  if  $H'(0)<0$,   combining with the fact that  $H(0)=0$,   there exists a unique $d_\lambda< x_p$  such that $H(d_\lambda)=0$; if $H'(0)\geq 0$, we have   $H(x)\geq0$  for any $x\in[0,x_p].$ 
  When $H(x_p)\leq0$,  we have   $H(x)\leq0$  for any $x\in[0,x_p].$
\end{itemize}
Define  $$ d_\lambda =\sup\{x:~1-S_X(X)-\lambda h_1(S_X(x))\leq 0 ~~\text{and} ~~0\leq x\leq x_p\},$$ 
then  we have   $I(x)=I_{d,x_p}(x):=\int_0^x q(t)\d t =x\wedge d_\lambda+  (x-x_p)_+$. In particular,  if $d_\lambda=x_p$,    then $I(x)=x$.  
 That is,  for every $s$,  there exists an $I_{d,x_p}(x)= x\wedge d+(x-x_p)_+$ that does better than any $I\in\mathcal{I}$. 

Next, we  aim to find the optimal $d$ for  problem  \eqref{eq:prob3} when  the insurance contract is given by $I_{d,x_p}$ for some $0\leq d \leq x_p$, that is,  
\begin{align}\label{eq:F2}\min _{d \in [0,x_p]} F(d):=  \int _ {0 }^{ d } (1-S_X(x))   \d x  - \int _ {x_p}^{M } S_X(x)   \d x  + g\left( \int _{ d } ^{x_p}h_1(S_X(x))  \d x\right)+ \E[X].\end{align} 
By the first-order condition, we have
\begin{align*}F'(d)&= -g'\left(  \int _{ d } ^{x_p}h_1(S_X(x))  \d x\right) h_1(S_X( d ))     +(1-S_X(d)).\end{align*} 
We have $F'(0)=0$ and $F'(x_p)=-g'(0)h_1(1-p)+p.$
It is straightforward to check that $$F''(d)= g''\left(  \int _{ d } ^{x_p}h_1(S_X(x))  \d x\right) h^2_1(S_X( d ))  -g'\left(  \int _{ d } ^{x_p}h_1(S_X(x))  \d x\right) h'_1(S_X( d )) S'_X(d)    -S'_X(d)).$$
Since  $g$ is convex, $S_X(d)$ decreases in $d$ and $h$ is concave,  $F''$  has at most one intersection with the x-axis.  Define  $$d^* =\sup\{x:~1- S_X(x)-g'\left(  \int _{x }^{x_p} h_1(S_X(t))  \d t\right) h_1(S_X(x))\leq0~~\text{and} ~~0\leq x\leq x_p\}.$$   Then  $d^*$  is the unique optimal solution to \eqref{eq:F2}.  
\end{proof}

So, if insurance premium is based on the VaR, the optimal indemnity is a dual truncated stop-loss indemnity. To be precise, the optimal indemnity provides full coverage for small losses up to a limit, and additionally for losses beyond another deductible that is based on $\VaR_p(X)$. This implies that the retained loss after insurer is bounded: $X-I^*(X)\leq \VaR_p(X)-d^*$.
We remark that  the optimal solution for $\Pi=\VaR_p$ with $p\in(0,1)$ is unique. This is because $S_X$ is strictly decreasing on $[0,M]$, therefore it holds that the set $\{x\in[0,M]|~ -S_X(x)-\lambda h(S_X(x))=0\}$ has Lebesgue measure zero. 

Again, we focus on $D=\mathrm{Gini}$ to illustrate the behavior of $d^*$ when the premium is based on the VaR. Since the behaviors under exponential distribution and uniform distribution are similar, we only give the results of uniform distribution in the following examples.
\begin{example}\label{exm:3} Let $h(t)=t-t^2$ with $t\in[0,1]$ and $g(x)=\alpha x+ \beta x^2$ with    $\alpha\geq0$.  Take $\theta=0.2$ and $p=0.9$.
If  $X\sim U[0,b]$, we have $x_p=p b$. Then $d^*$  in \eqref{eq:d_VaR}  becomes
$$d^*=\sup\left\{x: \frac{x}{b}-\frac{bx-x^2}{b^2}\left(\alpha+\frac{\beta}{b^2}\left(b x_p^2-\frac {2}{3} x_p^3-bx^2+\frac{2}{3}x^3\right)\right) \leq 0, ~~\text{and} ~~0\leq x\leq x_p\right\}.$$
In Figure \ref{fig:3}, we display the threshold $d^*$ as a function of $b$ (left figure) and the optimal indemnities as a function of quantile $p$. Overall, we can see that the threshold $d^*$ is increasing in $b$, and strictly increasing whenever the threshold is strictly positive. Also, we can see that a larger threshold $d$ is associated with larger values of $\alpha$ and $\beta$, because a larger weight on the Gini-deviation means that the DM prefers to purchase more insurance.
\begin{figure}[htb!]
\centering
 \includegraphics[width=15cm]{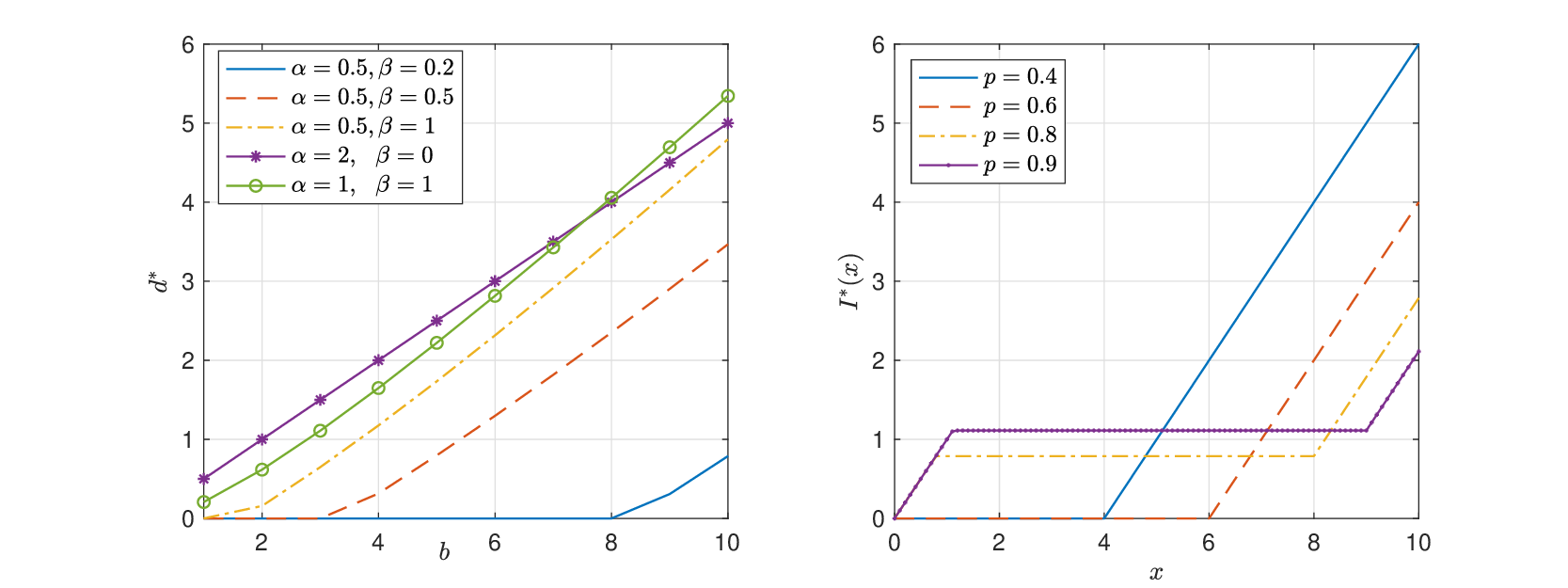} \captionsetup{font=small}
 \caption{ \small Optimal threshold $d^*$ with $p=0.8$  (left figure) and  optimal indemnity function $I^*$ with $b=10,~ \alpha=0.5,~\beta=0.2$  (right figure).}\label{fig:3}
\end{figure}

\end{example}
 \subsection{Expected Shortfall}
We next give the optimal results for $\Pi=\ES_p$ for $p\in(0,1)$. The following proposition shows that the optimal indemnity has a similar structure as for the case with $\Pi=\VaR_p$ (see Proposition \ref{prop:1}), but with a more complex selection of the deductible parameter (denoted as $d^*_{2}$ below) beyond which the indemnity provides full marginal coverage.  
\begin{proposition}\label{prop:2}
 Suppose that $D$ is given by \eqref{eq:D_h}, and $h_2(t)=\frac{t}{1-p} \wedge 1$ with $p\in(0,1)$, i.e., $\Pi_{h_2}(X)=\ES_p(X)$. The following statements hold:
 \begin{enumerate}[(i)]
     \item For every $I\in\mathcal I$,
 we can construct a  dual truncated stop-loss 
 insurance treaty   $I_{d_1,d_2}(x)=x\wedge d_2+  (x-d_1)_+$ for some $0\leq d_2\leq x_p<d_1\leq M$ such that $\mathrm{MD}^D_g (T_{I_{d_1,d_2}})\leq \mathrm{MD}^D_g (T_{I})$.  
 Further, $  I_{d_1^*, d_2^*}$ with   
 \begin{equation}\label{eq:d1_ES}
     d^*_{1} =\sup\left\{x:~ g'\left(  \int _{  d^*_{2} } ^{ x}h_1(S_X(t))  \d t\right)  h_1(S_X(x))-\frac{p}{1-p}S_X(x)\leq 0~\text{and}~x_p<x<M\right\},
\end{equation} and\begin{equation}\label{eq:d2_ES}d^*_{2} =\sup\left\{x:~1-S_X(x)- g'\left(  \int _{ x} ^{ d_{1}}h_1(S_X(t))  \d t\right) h_1(S_X(x))\leq 0~~\text{and} ~0\leq x\leq x_p\right\},\end{equation}
  is a solution to problem \eqref{eq:prob3}.
\item  If  $h''(0)<0$, the optimal solution  to problem \eqref{eq:prob3}   is unique on  $[0,M]$,  i.e., we have $I_{d_1^*,d_2^*}=\arg\min _{I \in \mathcal{I}} \mathrm{MD}^D_g (T_{I})$.
 \end{enumerate}
\end{proposition}

\begin{proof} 
The steps are the similar as Proposition \ref{prop:1},   and   $q$ in \eqref{eq:general}  minimizes \eqref{eq:f5} when $h_2(t)=\frac{t}{1-p} \wedge 1$ since it holds for a general $h_2\in \widetilde H_1$. Again, let $H(x)=h_2(S_X(x))-S_X(x)- \lambda h_1(S_X(x)).$ 
 \begin{itemize}
 \item[(i)] For $t<1-p$, or equivalently, $x>x_p$, we  have $H(x)=\frac{p}{1-p}S_X(x)- \lambda h_1(S_X(x))$ and  thus $H(x_p)= p-\lambda h_1(1-p)$. When $H(x_p)>0$,  if $H'(M)=(\frac{p}{1-p}-\lambda h'_1(0))S'_X(M)>0$, then there exists a unique $d_{1 \lambda}$ such that $H(x)>0$ for $x_p<x<d_{1 \lambda}$, and   $H(x)<0$ for $d_{1 \lambda}<x<M$;   if $H'(M)\leq 0$, then $H(x)\geq 0$ for any $x\in(x_p,M]$.  When   $ H(x_p)\leq 0$, then we have $H_1(x)\leq0$  for any $x_p<x<M$. 

\item[(ii)] For  $t\geq1-p$, or equivalently,  $x\leq x_p$, the analysis is similar to the case of VaR.  
\end{itemize}
  Define  $$\underline d_{1\lambda} =\sup\left\{x:~\frac{p}{1-p}S_X(x)- \lambda h_1(S_X(x))>0, ~\text{and}~x_p<x<M\right\},$$ 
  $$\overline d_{1\lambda} =\sup\left\{x:~\frac{p}{1-p}S_X(x)- \lambda h_1(S_X(x))\geq 0, ~\text{and}~x_p<x<M\right\},$$
  and
$$d_{2\lambda} =\sup\left\{x:~1-S_X(X)-\lambda h_1(S_X(x))\leq 0 ~~\text{and} ~0\leq x\leq x_p\right\}.$$  
 It is clear that $0\leq d_{2\lambda} \leq x_p<\underline d_{1\lambda}\leq \overline  d_{1\lambda} $. Thus, similar to Theorem \ref{thm:1},  we can select the function $c$ to be of the form $c(x)=1_{\{x>d_{1\lambda}\}}$ for some $d_{1\lambda}\in[\underline d_{1\lambda},\overline d_{1\lambda}]$, and   $I(x)=I_{d_{1\lambda},d_{2\lambda}}(x):=\int_0^x q(t)\d t =x\wedge d_{2\lambda}+  (x-d_{1\lambda})_+$.  Now, $\lambda$ is such that $D_h(X-I_{\underline d_{1\lambda},d_{2\lambda}}(X))\geq s$ and $D_h(X-I_{\overline d_{1\lambda},d_{2\lambda}}(X))\leq s$, and since $D_h(X-I_{d_{1\lambda},d_{2\lambda}}(X))$ is increasing in $d_{1\lambda}$,  there exists $d_{1\lambda}\in[\underline d_{1\lambda},\overline d_{1\lambda}]$ such that 
$$s=D_h(X-I_{d_{1\lambda},d_{2\lambda}}(X))=  \int _{d_{2\lambda}}^{d_{1\lambda}} h(S_X(x))  \d x.$$ That is,  for every $s$,  there exists an $I_{d_1,d_2}(x)= x\wedge d_2+(x-d_1)_+$ that does better than any $I\in\mathcal{I}$.

Next, we show that for any $I^*$ that solves  
\eqref{eq:prob},  there exists an $I_{d_1,d_2}(x)=x\wedge d_2+(x-d_1)_+$ such that $\mathrm{MD}^D_g(T_{I^*})=\mathrm{MD}^D_g(T_{I_{d_1,d_2}})$.  We fix that $s=D_h(X-I^*(X))$. By the above steps, for any given $s$,   we can always construct an insurance treaty $I_{d_1,d_2}(x)=x\wedge d_2+(x-d_1)_+$ for some $0 \leq d \leq M$ such that $ \mathrm{MD}^D_g(T_{I_{d_1,d_2}})\leq \mathrm{MD}^D_g(T_{I^*})$.      Since $I^*$  is optimal, then we have $\mathrm{MD}^D_g(T_{I^*})=\mathrm{MD}^D_g(T_{I_{d_1,d_2}})$.

Finally,  we  aim to find the optimal $d_1, d_2$ for the problem  \eqref{eq:prob3} when  the insurance contract is given by $I_{d_1,d_2}$ for some $0\leq d_2\leq x_p<d_1\leq M$, that is,  
 \begin{align*}\min _{0\leq d_2\leq x_p<d_1\leq M} F(d_1,d_2):=  \int _ {0 }^{ d_{2} } (1-S_X(x))   \d x  + \int _ {\d_{1} }^{M }\left(\frac{p}{1-p}S_X(x)\right)   \d x  + g\left( \int _{  d_{2} } ^{ d_{1}}h_1(S_X(x))  \d x\right)+ \E[X].\end{align*} 
By the first-order condition, we have
\begin{align*}\frac{\partial F(d_1, d_2)}{\partial d_1}&=g'\left(  \int _{  d_{2} } ^{ d_{1}}h_1(S_X(x))  \d x\right) h_1(S_X( d_{1} ))-\frac{p}{1-p}S_X(d_1).\end{align*} 
It is obvious that $$\frac{\partial F(d_1, d_2)}{\partial d_1}\Big|_{d_1=x_p}= g'\left(  \int _{  d_{2} } ^{x_p}h_1(S_X(x))  \d x\right) h_1(1-p)-p,~~~~\frac{\partial F(d_1, d_2)}{\partial d_1}\Big|_{d_1=M}=0.$$  Moreover, $$\frac{\partial F^2(d_1, d_2)}{\partial d^2_1}= g''\left(  \int _{  d_{2} } ^{ d_{1}}h_1(S_X(x))  \d x\right) h^2_1(S_X( d_{1} ))+g'\left(  \int _{  d_{2} } ^{ d_{1}}h_1(S_X(x))  \d x\right) h'_1(S_X( d_{1} ))S'_X( d_{1} )-\frac{p}{1-p}S'_X(d_1).$$
It is easy to check that $\frac{\partial F^2(d_1, d_2)}{\partial d^2_1}$ has  at most one intersection point with the x-axis, then we have 
 $$d^*_{1} =\sup\left\{x:~ g'\left(  \int _{  d^*_{2} } ^{ x}h_1(S_X(t))  \d t\right)  h_1(S_X(x))-\frac{p}{1-p}S_X(x)\leq 0~\text{and}~x_p<x<M\right\}.$$
Similarly,  we have  \begin{align*}\frac{\partial F(d_1, d_2)}{\partial d_2}&=-g'\left(  \int _{  d_{2} } ^{ d_{1}}h_1(S_X(x))  \d x\right) h_1(S_X( d_{2} ))+1-S_X(d_2),\end{align*} 
and 
\begin{align*}\frac{\partial F^2(d_1, d_2)}{\partial d^2_2}&=g''\left(  \int _{  d_{2} } ^{ d_{1}}h_1(S_X(x))  \d x\right) h^2_1(S_X( d_{2} ))-g'\left(  \int _{  d_{2} } ^{ d_{1}}h_1(S_X(x))  \d x\right) h'_1(S_X( d_{2} ))S'_X( d_{2})-S'_X(d_2).\end{align*}
By the similar arguments, we have 
  $$d^*_{2} =\sup\left\{x:~1-S_X(x)- g'\left(  \int _{ x} ^{ d^*_{1}}h_1(S_X(t))  \d t\right) h_1(S_X(x))\leq 0~~\text{and} ~0\leq x\leq x_p\right\}.$$
 This concludes the proof of (i). 
The proof of (ii) is similar to the one for  Theorem \ref{thm:1} (ii).

\end{proof} 

\begin{example}\label{exm:4} Let $h(t)=t-t^2$ with $t\in[0,1]$ and $g(x)=\alpha x+ \beta x^2$ with    $\alpha\geq0$ and $\beta\geq0$.  Take $\theta=0.2$.
If  $X\sim U[0,b]$, we have $x_p=p b$. Then $d^*_1$ and $d_2^*$ in \eqref{eq:d1_ES}  and \eqref{eq:d2_ES} become
     $$d^*_{1} =\sup\left\{x:~ \frac{bx-x^2}{b^2}\left(\alpha+\frac{\beta}{b^2}\left(b x^2-\frac {2}{3} x^3-bd_2^2+\frac{2}{3}d_2^3\right)\right)-\frac{p(b-x)}{(1-p)b} \leq 0~\text{and}~x_p<x<M\right\},$$  and 
  $$d^*_{2} =\sup\left\{x:~ \frac{x}{b}-\frac{bx-x^2}{b^2}\left(\alpha+\frac{\beta}{b^2}\left(b d_1^2-\frac {2}{3} d_1^3-bx^2+\frac{2}{3}x^3\right)\right) \leq 0~~\text{and} ~0\leq x\leq x_p\right\}.$$
In Figure \ref{fig:4}, we display these two thresholds as a function of $b$ for two sets of parameters. We can see that $d_2^*=0$ for all $b\in[0,10]$ in the left figure, which suggests that the optimal indemnity is of a stop-loss form. For larger values of $b$, this observation does not hold true in the middle figure. In both figures, the parameters $d_1^*$ and $d_2^*$ are increasing in $b$, and strictly increasing whenever $d_2^*$ is strictly positive. Moreover, the right figure shows three optimal indemnity functions for three different choices of $p$. Interestingly, we can see that for larger values of $p$, the second parameter $d_1^*$ is larger, and thus the indemnity functions provide less coverage in the right tail. 
  \begin{figure}[htb!]
\centering
 \includegraphics[width=18cm]{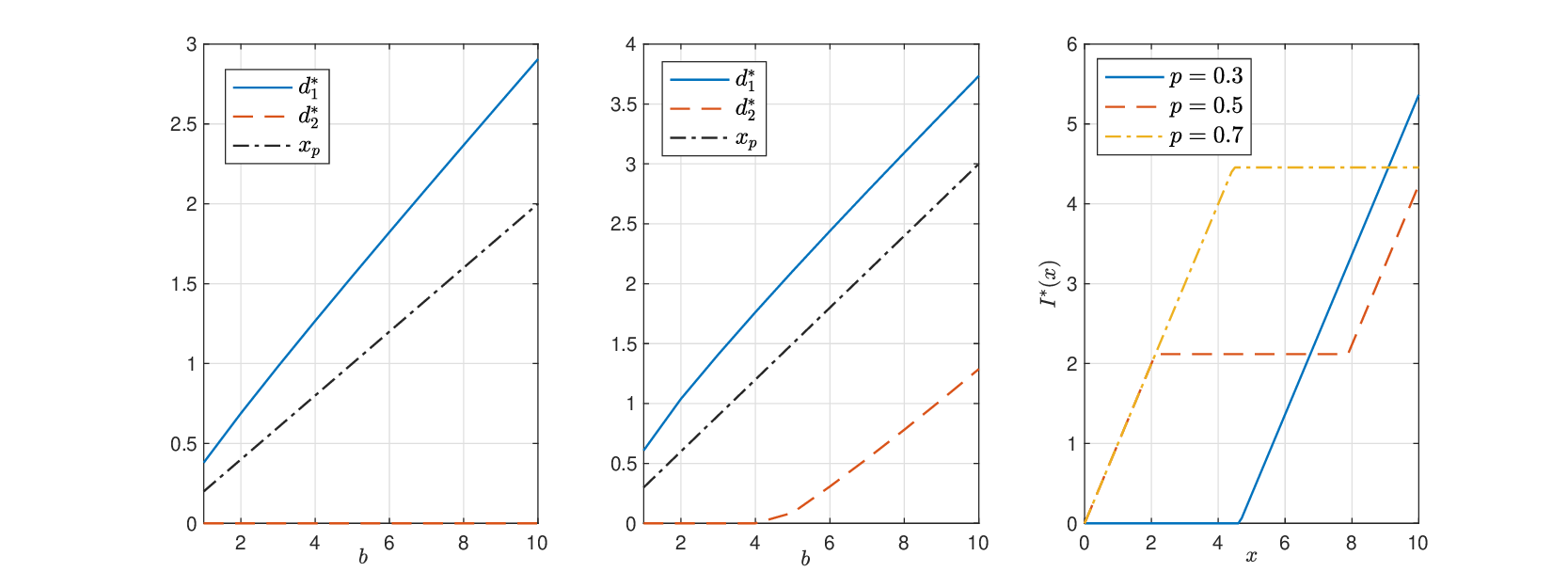} \captionsetup{font=small}
 \caption{ \small  Optimal parameters $d_1^*$ and $d_2^*$ corresponding to Example \ref{exm:4} for the cases $p=0.2, \alpha=0.5, \beta=0.3$  (left figure),  and $p=0.3, \alpha=0.7, \beta=0.5$ (middle figure). The right figure shows the optimal indemnity for three choices of the parameter $p$, with  $b=10,~ \alpha=0.5,~\beta=0.3$.}\label{fig:4}
\end{figure}
\end{example} 


\section{The budget constraint problem}\label{sec:5}
In this section, we assume that  the insurer faces a fixed budget to purchase insurance. This yields the following constraint:
\begin{equation}\label{eq:cons}
\Pi(I(X)) \leq \overline \Pi, \quad \text { for some budget threshold } \overline \Pi>0 .\end{equation}
We refer to the minimization problem \eqref{eq:prob} subject to \eqref{eq:cons} as the budget constraint problem.  For simplicity, we focus in this section only on the cases under which we showed uniqueness of the optimal solution in Sections \ref{sec:EVPP} and \ref{sec:DPP}.  

 Assume that  an unconstrained optimal solution $I^*$ has premium  equal to $\Pi_0=\Pi(I^*(X))$. To avoid redundant 
arguments, we assume  $\overline{\Pi}< \Pi_0$,  that is, $\overline{\Pi}$ is no larger than the minimal premium for optimal solutions without budget constraint. This means that the optimal solution to the unconstrained problem is no longer feasible in the constrained problem.

\begin{proposition}When  $\Pi$ is calculated by the expected value premium principle in  \eqref{eq:ev} or the distortion premium principle in \eqref{eq:dis_pre},  the constraint \eqref{eq:cons} is binding  to \eqref{eq:prob} for  $\overline \Pi <\Pi_0$. 
\end{proposition}
\begin{proof}Suppose \eqref{eq:prob} with \eqref{eq:cons} admits a solution $\widetilde{I}$ for which the constraint \eqref{eq:cons} is slack.  Note that
$$
\mathrm{MD}^D_g(X-I^*(X)+\Pi(I^*(X)))<\mathrm{MD}^D_g(X-\widetilde I(X)+\Pi(\widetilde I(X))).
$$
There exists $\lambda \in(0,1)$ such that
$
\Pi(I(X))=\lambda\Pi(\widetilde{I}(X))+(1-\lambda) \Pi(I^*(X))=\overline \Pi,
$ where $I=\lambda \widetilde I +(1-\lambda) I^* $ due to the fact that both the expected premium principle and the distortion premium principles are comonotonic additive.  Since $\mathrm{MD}^D_g$ is convex, we have $$\begin{aligned}\mathrm{MD}^D_g(X- I(X)+\Pi(I(X)))&= \mathrm{MD}^D_g(\lambda (X- \widetilde I(X)+\Pi(\widetilde I(X)))+(1-\lambda)(X-  I^*(X)+\Pi(I^*(X)))) \\&\leq \lambda \mathrm{MD}^D_g(X-\widetilde I(X)+\Pi(\widetilde I(X)))  +(1-\lambda)  \mathrm{MD}^D_g(X-I^*(X)+\Pi(I^*(X))) \\& < \mathrm{MD}^D_g(X-\widetilde I(X)+\Pi(\widetilde I(X))), \end{aligned},$$which contradicts the optimality of $\widetilde I$.  Thus, the constraint \eqref{eq:cons}  should be binding  to \eqref{eq:prob}.
\end{proof}

\begin{theorem}\label{thm:3} Suppose $\overline\Pi<\Pi_0$, and one of the following holds:\begin{itemize}\item $D=D_h$ with  $h''(0)<0$ as given by \eqref{eq:D_h}, or 
\item $D=\mathrm{SD}$  and $g(x)=\alpha x+\beta x^2$ with $\alpha\geq0$ and $\beta\geq0$.\end{itemize} Then,  the optimal indemnity $\widetilde I^* \in \mathcal{I}$ for \eqref{eq:prob} with constraint \eqref{eq:cons}  is given by $$\widetilde I^*(x)=(x-\widetilde d^*)_+,$$ where   
$\widetilde d^*$ is the solution to $\Pi((X-\widetilde d^*)_+)=\overline \Pi$.

\end{theorem} \begin{proof} \underline{Case 1: $D=D_h$}. We fix $D_h(X-I(X))=s\in [0,D_h(X)]$ and solve \eqref{eq:prob} subject to  constraint \eqref{eq:cons}.  We  translate the constrained minimization problem  to a non-constrained problem by using the Lagrangian multiplier
method. Consider the following minimization problem 
$$ \min _{I \in \mathcal{I}} \widetilde   f(I):=g(s)+\theta \E[I(X)]+ \E[X]+\lambda_1 (D_h(X-I(X)))-s)+\lambda_2 ((1+\theta)\E[I(X)]-\overline \Pi)$$ with $\lambda_1$ and $\lambda_2$  being the KKT multipliers. By  similar arguments as in the proof of Theorem \ref{thm:1},  we can  write
\begin{equation}\label{eq:f2}\begin{aligned}\widetilde f(I)= & \int _0^M( \theta S_X(x)+\lambda_2(1+\theta) S_X(x) -\lambda_1 h(S_X(x)))  q(x) \d x \\&  + \lambda_1 D_h(X)+ g(s)+  \E[X]-\lambda_1 s-\lambda_2 \overline \Pi.\end{aligned}\end{equation}
Let  $$H(x)= \theta S_X(x) +\lambda_2(1+\theta)S_X(x) -\lambda_1 h(S_X(x)).$$  
For any $\lambda_1, \lambda_2\in[0,\infty)$,  it is clear that  $H(0)=\theta+\lambda_2(1+\theta)>0$, $\lim_{x\to M}H(x)=0$, and  $$  H'(x)= (\theta+\lambda_2(1+\theta) -\lambda_1 h'(S_X(x)))S'_X(x).$$  Since $h$ is a concave function with $h(0)=h(1)=0$, 
  if $\theta+\lambda_2(1+\theta)-\lambda_1 h'(0)<0$,  there exists a unique   $ d _{\lambda_1,\lambda_2}\in[0,M)$ such that   $  H(x)<0$ for $x\in ( d_{\lambda_1,\lambda_2}, M)$ and $  H(x)> 0$ for  $x\in [0,  d_{\lambda_1,\lambda_2} )$. 
Thus, if $\theta+\lambda_2(1+\theta)-\lambda_1 h'(0)<0$, then  the following $\widetilde q $ will minimize \eqref{eq:f2}
$$\widetilde   q(x)=\left\{\begin{aligned}&0,~~~&\text{if } H(x)>0~~ (i.e.,  ~x< d_{\lambda_1,\lambda_2} ),\\&1,~~~&\text{if } H(x)<0 ~~(i.e., ~x> d_{\lambda_1,\lambda_2} ),\\& c,~~~&otherwise,\end{aligned}\right.$$
where $c$ could be any $[0,1]$-valued constant on $ H(x)=0$ (i.e., $x= d_{\lambda_1,\lambda_2}$). 
On the other hand, if $\theta+\lambda_2(1+\theta)-\lambda_1 h'(0)\geq0$, $  H'(x)\leq 0$ for all $x\geq 0$, which implies $ H(x)\geq 0$ for all $x\geq 0$. In this case, $d_{\lambda_1,\lambda_2}=M.$ Then we have $I(x)=I_{d_{\lambda_1,\lambda_2}}(X):=\int_0^x q(t)\d t =(x-d_{\lambda_1,\lambda_2})_+$.

 Next, we  aim to find the optimal $d$ for problem  \eqref{eq:prob} subject to \eqref{eq:cons} when  the insurance contract is given by $I_d$ for some $d\in[0,M]$, that is, 
 \begin{align}\label{eq:wF1} \min _{d \in [0,M]} \widetilde   F(d)= & \int _ { d }^M( \theta S_X(x)+\lambda_2(1+\theta) S_X(x)  \d x  + g\left( \int _0^{ d  } h(S_X(x))  \d x\right)+  \E[X]-\lambda_2 \overline \Pi .\end{align} 
By the first-order condition, we have  
\begin{align*}\widetilde   F'(d)&=g'\left( \int _0^{ d  } h(S_X(x))  \d x\right) h(S_X(d))-(\theta S_X(d)+\lambda_2(1+\theta) S_X(d)).\end{align*} 
Assume that there exists a constant $\lambda^*_2 \geq 0$ such that $d_{\lambda^*_2}$ solves problem \eqref{eq:wF1} for $\lambda_2=\lambda^*_2$ and $\int _{d_{\lambda^*_2} }^M  (1+\theta) S_X(x) \d x=\overline \Pi$.  Then, we can show $\widetilde d^* =\widetilde  d_{\lambda^*_2}$ solves problem \eqref{eq:prob} subject to the constraint \eqref{eq:cons}. 
 We denote the optimal value of problem \eqref{eq:prob} with constraint \eqref{eq:cons} by $V(\overline\Pi)$. Then, it follows that
$$\begin{aligned}
V(\overline\Pi) & =\sup _{\substack{d \in [0,M] \\
\int _{d }^M  (1+\theta) S_X(x) \d x \leq \overline \Pi}} \mathrm {MD}_g^D(T_I) 
\leq \sup _{\substack{d \in [0,M] \\
\int _{d }^M  (1+\theta) S_X(x) \d x \leq \overline \Pi}} \left\{\mathrm {MD}_g^D(T_I)-\lambda^*_2\left( \int _{d }^M  (1+\theta) S_X(x) \d x-\overline \Pi \right)\right\}  \\
& \leq \sup _{d \in [0,M]} \left\{\mathrm {MD}_g^D(T_I)-\lambda^*_2\left( \int _{d }^M  (1+\theta) S_X(x) \d x-\overline \Pi \right)\right\}   =\mathrm {MD}_g^D(T_{I_{\widetilde d_{\lambda_2^*}}}) \leq V(\overline\Pi).
\end{aligned}$$ The last inequality is because $I_{\widetilde d_{\lambda_2^*}}$ is feasible to problem \eqref{eq:prob}  without the constraint. 
Hence, $\widetilde d^*=\widetilde d_{\lambda^*_2}$ solves problem problem \eqref{eq:prob} subject to \eqref{eq:cons}.
Thus,  we have  $\Pi((X-\widetilde d^*)_+)=\overline \Pi$. In this case, $\lambda_2$ can be solved by 
\begin{equation*}\label{eq:tilde_d} \lambda^*_2=\inf\left\{\lambda_2:~g'\left( \int _0^{ \widetilde d^* } h(S_X(x))  \d x\right) h(S_X(x))-\theta S_X(x)+\lambda_2(1+\theta)S_X(x)\leq 0,~\text{and}~\lambda_2\geq0\right\}.\end{equation*}

\underline{Case 2: $D=\mathrm{SD}$.} Let   $g(x)=\alpha x+\beta x^2$ with $\alpha\geq 0$ and $\beta\geq 0$.  With the budget constraint, we first consider the following minimization problem \begin{equation}\begin{aligned}  \label{eq:wf2} \inf_{0\leq d< M} \widetilde f(d):= &g(\mathrm{SD}(X\wedge d))+ \E[X]+\theta \E[(X-d)_+]+\lambda ((1+\theta) \E[(X-d)_+]-\overline \Pi)\\ =&g\left( \left (w_2(d)-w^2_1(d)\right)^{1/2} \right)+\E[ X] +(\theta +\lambda(1+\theta) )\int_d^M  S_X(x)\d x- \lambda \overline \Pi.\end{aligned}\end{equation}
 We only need to replace $\theta$ in Theorem \ref{thm:2}  with $\theta +\lambda(1+\theta)$. 
 By the first order condition,  we have 
\begin{align*}\widetilde  f'(d)=& \frac{1}{2 \sqrt {w(d)}}g' ( \sqrt{w(d)} )w'(d)-(\theta +\lambda(1+\theta) )S_X(d)\\=& S_X(d)\left(   \frac{g' ( \sqrt{ w(d)}) }{\sqrt {w(d)} } (d-w_1(d)) -(\theta +\lambda(1+\theta) )\right). \end{align*}
 For $g(x)=\alpha x+\beta x^2$ with $\alpha\geq0$ and  $\beta\geq0$,  it becomes that  \begin{align*}  \widetilde  f'(d) = S_X(x)\left(\alpha \sqrt {\frac{(d-w_1(d))^2}{w_2(d)-w^2_1(d)} }+2\beta  (d-w_1(d)) -(\theta +\lambda(1+\theta) )\right).\end{align*}
Again, assume that there exists a constant $\lambda^* \geq 0$ such that $d_{\lambda^*}$ solves problem \eqref{eq:wf2} for $\lambda=\lambda^*$ and $\int _{d_{\lambda^*} }^M  (1+\theta) S_X(x) \d x=\overline \Pi$.  Then, we aim to show $\widetilde d^* =\widetilde  d_{\lambda^*}$ solves problem  \eqref{eq:prob} with constraint \eqref{eq:cons}. The process  is  similar to the first part, and   we have
 $\Pi((X-\widetilde d^*)_+)=\overline \Pi$. In this case, $\lambda$ can be solved by 
  $$  \lambda^*=\inf \left \{\lambda:    \alpha \sqrt {\frac{(x-w_1(\widetilde d^*))^2}{w_2(\widetilde d^*)+w^2_1(\widetilde d^*)} }-2\beta  (\widetilde d^*-w_1(\widetilde d^*))- (\theta +\lambda(1+\theta) ) \leq 0,~\text{and}~\lambda\geq0\right\},$$ which yields the results.  
\end{proof}

\medskip
Recall Example \ref{exm:1} in Section \ref{sec:3.1} where $D=\mathrm{Gini}$, we further assume that DM has a budget $\overline \Pi$ on his purchasing of insurance. 
\begin{example} Let $h(t)=t-t^2$ with $t\in[0,1]$ and $g(x)=\alpha x+ \beta x^2$ with    $\alpha=0.5$ and $\beta=0.7$. Based on Example \ref{exm:1}, we can compute that $d^*=2.39$ for $X\sim U[0,10]$. Since  $\theta=0.2$, we have $\Pi(I^*)=(1+\theta)\E[I^*(X)]=3.48$, and thus we   assume that $\overline \Pi<3.48$. Similarly, we can compute  $d^*=2.55$ for $X\sim \mathrm{exp}(0.1)$ and $\Pi(I^*)=(1+\theta)\E[I^*(X)]=9.73$; thus we assume that $\overline \Pi<9.73$. \begin{figure}[htb!]
\centering
 \includegraphics[width=15cm]{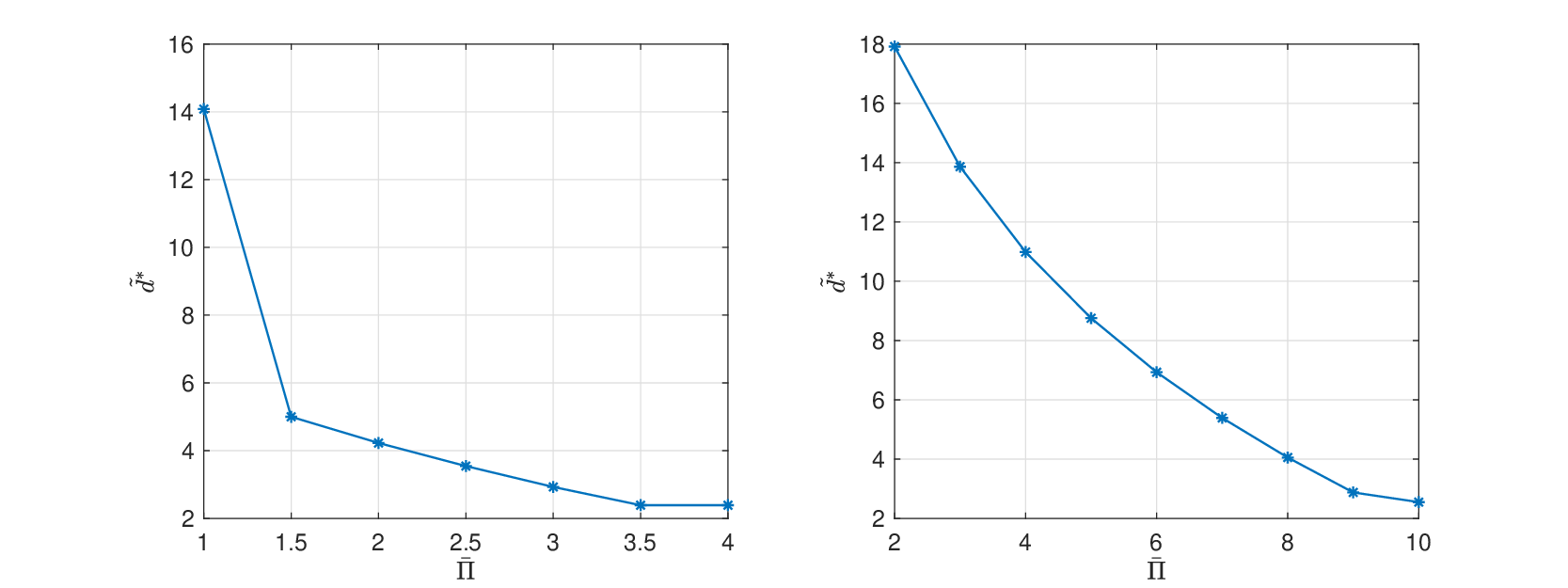} \captionsetup{font=small}
 \caption{ \small Optimal deductible $\widetilde d^*$ for the uniform distribution with $a=0$ (left figure) and exponential distribution with  $\mu=10$ (right figure).}\label{fig:5}
\end{figure} 

We can see from Figure \ref{fig:5} that the optimal deductible  increases as the constraint $\overline\Pi$ increases, which implies that the DM chooses to retain more claims if the premium  budget is relatively small. In particular, when the  budget is relatively larger, say $\overline\Pi>3.5$ in left panel of Figure \ref{fig:5} and $\overline\Pi>9.73$ in right panel of Figure \ref{fig:3}, the constraint is not binding. Thus, the optimal results are identical to those without constraint.
\end{example}

We next present the optimal indemnity function with a budget constraint when the premium is calculated by VaR or ES. We will show that the optimal indemnity remains a dual truncated stop-loss indemnity when we add the budget constraint, but the corresponding parameters are modified.  Since the proof is similar to Propositions \ref{prop:1}-\ref{prop:2} and Theorem \ref{thm:3}, we only present the major steps
that highlight the differences. Also, the proof is relatively lengthy, so we put it in Appendix \ref{sec:AppC}. 
\begin{theorem}
    \label{thm:4} Suppose $\overline\Pi<\Pi_0$ and $D=D_{h_1}$ with $h_1\in\mathcal{H}_0$ and $h''_1(0)<0$, when $\Pi=\VaR_p$ or $\ES_p$ for some $p\in[0,1)$, 
 the optimal indemnity $\widetilde I^* \in \mathcal{I}$ for \eqref{eq:prob} with constraint $\Pi(\widetilde I^*(X))\leq\overline \Pi$  is given by $$\widetilde I^*(x) =(x-\widetilde  d^*_1)_+\wedge (\widetilde d^*_2-\widetilde d^*_1)+  (x-\widetilde d^*_3)_+.$$  
Here,  $\widetilde  d^*_1$,  $\widetilde  d^*_2$ and $\widetilde d^*_3$ can be derived by solving 
\begin{align*} \widetilde  d^*_1 &=\inf\{x:~(1+\lambda_2-S_X(X))-\lambda_1h_1(S_X(x))\leq 0, ~~\text{and} ~~0\leq x\leq x_p\},\\ \widetilde  d^*_2 &=\sup \{x:~(1+\lambda_2-S_X(X))-\lambda_1h_1(S_X(x))\leq 0, ~~\text{and} ~~\widetilde  d^*_1\leq x\leq x_p\},\\ \widetilde  d^*_3 &=\left\{\begin{array}{ll} x_p &\text{ if }\Pi=\VaR_p,\\ \sup \left\{x:~\frac{p+\lambda_2}{1-p}S_X(x)-\lambda_1h_1(S_X(x))\leq 0, ~~\text{and} ~~x_p<x<M\right\} &\text{ if }\Pi=\ES_p,\end{array}\right.\end{align*}    where $\lambda_1$ and $\lambda_2$ are determined by $$\lambda_1 =g'\left(\int _{ 0 }^{\widetilde  d^*_1} h_1(S_X(x))  \d x+ \int _{ \widetilde  d^*_2}^{\widetilde  d^*_3} h_1(S_X(x))  \d x\right),$$ $$\lambda_2= \inf \left\{ \lambda_2: \widetilde  d^*_2-\widetilde  d^*_1-\overline \Pi \leq 0,~\text{and}~\lambda_2\geq0\right\}.$$ 
\end{theorem}

Note that for the VaR, the parameter $\widetilde  d^*_3$ does not change after we add the budget constraint. The reason is that increasing this parameter beyond $x_p$ reduces the coverage, but not the corresponding premium. 
Also note that if $\Pi$ is large enough, it will hold that  $\lambda_2=0$, and  then we have $d^*_1=0$. This thus recovers the structure of the indemnity function in the unconstrained case in Propositions \ref{prop:1}-\ref{prop:2}.
\section{Conclusion}\label{sec:conclusion}
This paper contributes to the field of optimal insurance contract theory by introducing and analyzing the use of mean-deviation measures as an objective for decision-makers. The findings highlight the desirability of stop-loss insurance indemnities and provide valuable insights into the optimal design of insurance contracts under different premium principles. Further research can build upon these results by exploring additional deviation measures and their implications for insurance contract optimization.

There are several possible extensions of the research presented in this paper. First, future research could explore the use of other deviation measures. Our focus in this paper is on convex signed Choquet integrals and the standard deviation. Second, the paper focuses on the case when the premium principle is either based on expected value, Value-at-Risk, or Expected Shortfall. Future research could investigate other premium principles and their implications on optimal insurance contract design. Finally, the paper only considers a single policyholder that is used to determine the premium charged by the insurer. Future research could examine the implications of multiple policyholders on optimal insurance contract design and explore the use of game theory in this context.

 \appendix
\section{Some background on risk measures}\label{sec:AppA}

In this appendix we collect some common terminology and results on risk measures, which are
briefly mentioned in the text of the paper, but not essential to the presentation of our main results.  All random variables are tacitly assumed to be in the space  $\X$. 

We next list some properties of risk measures. To do so, we first define \emph{comonotonicity}. A random vector $(Z_1,\dots,Z_n)$ is comonotonic
if there exists a random variable $Z$ and increasing functions $f_1,\dots,f_n$ on $\R$ such that $Z_i=f_i(Z)$ a.s.~for every $i=1,\dots,n$.  We define the following properties for a mapping $\rho: \mathcal{X} \rightarrow \mathbb{R}$: 
\begin{itemize}
\item[(A1)] (Law invariant)  $\rho(Y)=\rho(Z)$  for  all $Y,Z\in \X$  if  $Y$ and $Z$ follow the same distribution, 
\item[(A2)] (Cash invariance) $\rho (Y+c)= \rho (Y)+c$ for all  $c\in\R$,\item[{(A3)}] (Monotonicity)  $\rho(Y)\le \rho(Z)$ for all $Y,Z \in \mathcal X$ with $Y\le Z$,
\item[{(A4)}] (Convexity) $\rho(\lambda Y+(1-\lambda)Z)\le \lambda\rho(Y)+(1-\lambda)\rho(Z)$ for all  $Y,Z\in \X$ and $\lambda\in[0,1]$,
\item[(A5)] (Comonotonic additive)  $\rho(Y+Z)=\rho(Y)+\rho(Z)$ whenever $Y$ and $Z$ are comonotonic.
\end{itemize} Here, (A1) states that the risk value depends on the loss via its distribution. 
Using the standard terminology in \cite{FS16}, a risk measure is a \emph{monetary risk measure}  if it satisfies (A2) and  (A3),   it is a \emph{convex risk measure} if  it is monetary and further satisfies (A4), and it is a \emph{coherent risk measure}    if   it is monetary and further satisfies (D3) and (D4). Clearly,  (D3) together with  (D4) implies (A4). Thus, convex risk measures are more general than the coherent risk measures.

Below, we list some classic  convex signed Choquet integrals with $h\in\mathcal{H}_0,$ which are formulated on their respective
effective domains.  In fact, with some bounded assumptions of $\rho^c_h$ defined in \eqref{eq:quantile}, there    exists one-to-one correspondence between the deviation measures    and the  distortion risk measure   with the relation $\rho^0_h(X)=\rho^1_h(X)-\E[X]$.

\begin{enumerate}[(i)]
    \item The mean absolute deviation with $h(t)=t \wedge(1-t)$:
$$
\mathbb{E}[|X-\mathbb{E}[X]|], \quad X \in L^1.$$
\item  The Gini deviation with $h(t)=t-t^2$:
$$\frac{1}{2} \mathbb{E}\left[\left|X_1-X_2\right|\right], \quad X \in L^1, X_1, X_2, X ~\text{are ~iid}.$$
\item The range with $h(t)=\id_{\{0<t<1\}}$:
$$\operatorname{ess-sup}(X)-\operatorname{ess-inf}(X), \quad X \in L^{\infty}.$$
\item The inter-ES range with $h(t)=\frac{t}{1-\alpha} \wedge 1+\frac{\alpha-t}{1-\alpha} \wedge 0$: $$\mathrm{ES}_\alpha(X)+\mathrm{ES}_\alpha(-X), \quad \alpha \in(0,1), ~ X \in L^1.$$
\item The  ES deviation with $h(t)=\frac{\alpha t}{1-\alpha} \wedge (1-t)$:
$$\ES_\alpha-\E\quad \alpha \in(0,1),~ X \in L^1.$$
\end{enumerate}

\section{Monotonicity of  $\mathrm{MD}^D_g$}\label{sec:AppB}
The mapping $\mathrm{MD}^D_g$ in Definition \ref{def:1} is not necessarily monotonic, as defined as property (A2) in Appendix \ref{sec:AppA}. Thus, $\mathrm{MD}^D_g$ is generally not a monetary risk measure.  In fact, $\mathrm{MD}^D_g$ satisfies a weak  monotonicity which implies that $\mathrm{MD}^D_g(c_1)\leq\mathrm{MD}^D_g(c_2)$ if   $c_1\le c_2$  for any $c_1, c_2\in \R$. \cite{HWW23} characterized recently the mean-deviation measures which are
monotonic from a general mean-deviation model.  We define a \emph{mean-deviation model} below.
\begin{definition}[Mean-deviation model] Fix $p\in[1,\infty]$. A \emph{mean-deviation model} is a continuous functional  $U: L^p\rightarrow[0,\infty)$  defined as \begin{equation}\label{eq:V}
U(X)=V(\E[X], D(X))\end{equation}
where $V:\H\to (-\infty,\infty]$ with   $\H=\{(x,y)\in\R\times[0,\infty)\}$ such that
(i) $V(m, d)$ is strictly increasing in  $m$ for every $d$;
(ii) $V(m, d)$ is strictly increasing in $d$ for every $m$;
(ii) $V(m, 0)=m$ for every $m$ (normalization). \label{def:2}
\end{definition}
Theorem 1 of \cite{HWW23} showed that for a mean-deviation model $\rho=V(\E,D)$, $\rho$ is a monetary risk measure or further a consistent risk measure if and only if  $\rho= \mathrm{MD}^D_g$ where  $g:[0,\infty) \rightarrow \mathbb{R}$  is  a strictly increasing function  satisfying $1$-Lipschitz continuity and $g(0)=0$. Thus,  if we assume $g(x)=\alpha x+\beta x^2$ that is convex and  $D$ satisfies comonotonic additivity, to  make $\mathrm{MD}^D_g$ be monotonic on the relevant domain,   we must have    $g'(D(X-I(X)))=\alpha+2\beta (D(X)-D(I(X)) \leq 1$ for all $I\in\mathcal I$. 
Thus, since $D(I(X))\geq0$ by the non-negativity property (D2), we may assume that $$\{\alpha\geq0,~ \beta \geq0, (\alpha,\beta)\neq(0,0), ~~\text{and} ~ \alpha+2\beta D(X)\leq 1\}.$$ 
 Further, by  Proposition 4 of \cite{HWW23},  $\mathrm{MD}^D_g$ is a convex risk measure  if  and only if $$\mathrm{MD}^D_g(X)=c\E[(D(X)-Y)_+]+\E[X]$$   for some  non-negative random variable $Y\in L^1$ and some constant $0<c\leq1$.  In particular, $\mathrm{MD}^D_g$ is a  coherent risk measure  if $Y=0$. For instance, if we take $Y\sim U[0,b]$,  then we obtain the following convex risk measure:  
\begin{align*}\mathrm{MD}^D_g(X)=&c\E[(D(X)-Y)_+]+\E[X] =\frac{c}{b}\int_0^{D(X)\wedge b} (D(X)-y)\d y+\E[X]\\=&\frac{c}{b}(D(X)(D(X)\wedge b)-(D(X)\wedge b) ^2 )+\E[X].\end{align*}

\section{Proof of Theorem \ref{thm:4}}\label{sec:AppC}
\begin{proof}\underline{Case 1: $\Pi=\VaR_p$}. Along the similar lines in the proof of  Proposition \ref{prop:2} and Theorem  \ref{thm:3}, we consider the following minimization problem  \begin{equation}\label{eq:wf3}\begin{aligned}\min _{I \in \mathcal{I}} \widetilde   f(I):=&  \int _0^M((1+\lambda_2) h_2(S_X(x))-S_X(x)- \lambda_1 h_1(S_X(x)))  q(x) \d x  \\& + \lambda_1 D_{h_1}(X)+ g(s)+  \E[X]-\lambda_1 s -\lambda_2\overline\Pi,\end{aligned} \end{equation}
where  $\lambda_1,~\lambda_2\geq0$ are the KKT multipliers and   the following $q$ will minimize \eqref{eq:wf3}
\begin{equation}\label{eq:general1}q(x)=\left\{\begin{aligned}&0,~~~&\text{if }(1+\lambda_2)h_2(S_X(x))-S_X(x)- \lambda_1 h_1(S_X(x))>0,\\&1,~~~&\text{if } (1+\lambda_2) h_2(S_X(x))-S_X(x)- \lambda_1 h_1(S_X(x))<0 ,\\& c,~~~&otherwise,\end{aligned}\right.\end{equation}
where $c$ could be any $[0,1]$-valued constant on $(1+\lambda_2)h_2(S_X(x))-S_X(x)- \lambda_1 h_1(S_X(x))=0$.  
Define $$H(x)=(1+\lambda_2)h_2(S_X(x))-S_X(x)- \lambda_1h_1(S_X(x)).$$
Let  $h_2(t)=\id_{\{t>1-p\}}$ with $p\in(0,1)$. 
 \begin{itemize}
 \item[(i)] For  $t<1-p$, or equivalently, $x_p<x\leq M$, we always have $H(x)=-S_X(x)- \lambda_1 h_1(S_X(x))<0$ for $x_p<x< M$, which implies $q(x)=1$.  
\item[(ii)]For $t\geq1-p$, or equivalently,  $x\leq x_p$, we have $H(x)=1+\lambda_2-S_X(x)-\lambda_1 h_1(S_X(x)).$  Since $h_1$ is concave with $h_1(0)=h_1(1)=0$, $H'(x)=-S'_X(x)(1+\lambda_1 h'_1(S_X(x))$ and $H(x_p)=\lambda_2+p-\lambda_1 h(1-p)$, combining $H(0)=\lambda_2$,\footnote{Note that we have $H(0)=0$ in Proposition \ref{prop:1}, thus, there at most exists one zero such that $H(d)=0$.}  there at most exists two zeros   $d_{i, \lambda_1,\lambda_2}< x_p~(i=1,2)$  such that $H(d_{i, \lambda_1,\lambda_2})=0$. 
\end{itemize}
Define  \begin{equation*}d_{1,\lambda_1,\lambda_2} =\inf\{x:~1+\lambda_2-S_X(x)-\lambda_1 h_1(S_X(x))\leq 0 ~~\text{and} ~~0\leq x\leq x_p\},\end{equation*} and  \begin{equation*}d_{2,\lambda_1,\lambda_2} =\sup\{x:~1+\lambda_2-S_X(x)-\lambda_1 h_1(S_X(x))\leq 0 ~~\text{and} ~~d_{1,\lambda_1,\lambda_2}\leq x\leq x_p\}.\end{equation*}  
Then we  have   $I(x)=I_{d_{1,\lambda_1,\lambda_2},d_{2,\lambda_1,\lambda_2},x_p}(X):=\int_0^x q(t)\d t =(x-d_{1,\lambda_1,\lambda_2})_+\wedge (d_{2,\lambda_1,\lambda_2}-d_{1,\lambda_1,\lambda_2})+  (x-x_p)_+$. 
That is,  for every $s$,  there exists an $I_{d_1,d_2,x_p}(x)=(x-d_{1})_+\wedge (d_{2}-d_{1})+  (x-x_p)_+$ that does better than any $I\in\mathcal{I}$.   Next, we  aim to find the optimal $d_1$ and $d_2$ for problem  \eqref{eq:wf3} subject to \eqref{eq:cons} when  the insurance contract is given by $I_{d_1,d_2,x_p}$ for some $d_1, d_2\in[0,x_p]$, that is, 
\begin{equation} \begin{aligned}\label{eq:wF3} \min _{d_1, d_2 \in [0,x_p]} \widetilde   F(d_1,d_2):= & \int _ {d_{1} }^{d_{2}} (1+\lambda_2-S_X(x))   \d x  - \int _ {x_p}^{M } S_X(x)   \d x  \\&+ g\left(\int _{ 0 }^{d_{1}} h_1(S_X(x))  \d x+ \int _{ d_{2} }^{x_p} h_1(S_X(x))  \d x\right)+ \E[X] -\lambda_2\overline\Pi. \end{aligned}\end{equation}
 By the first-order condition, we have  
\begin{align*}\frac{\partial\widetilde   F(d_1,d_2)}{\partial d_1}&=g'\left(\int _{ 0 }^{d_{1}} h_1(S_X(x))  \d x+ \int _{ d_{2} }^{x_p} h_1(S_X(x))  \d x\right) h_1(S_X(d_1))-(1+\lambda_2- S_X(d_1)),\end{align*} and \begin{align*}\frac{\partial\widetilde   F(d_1,d_2)}{\partial d_2}&=-g'\left(\int _{ 0 }^{d_{1}} h_1(S_X(x))  \d x+ \int _{ d_{2} }^{x_p} h_1(S_X(x))  \d x\right) h_1(S_X(d_2))+(1+\lambda_2- S_X(d_2)).\end{align*} 
Next, assume that there exists a constant $\lambda^*_2 \geq 0$ such that $d_{1\lambda^*_2}$ and $ d_{2\lambda^*_2}$ solves problem \eqref{eq:wF3} for $\lambda_2=\lambda^*_2$ and $d_{2\lambda^*_2}-d_{1\lambda^*_2}=\overline \Pi$.  Then, we aim to show $\widetilde d^*_1 =\widetilde  d_{1\lambda^*_2}$ and  $\widetilde d^*_2 =\widetilde  d_{2\lambda^*_2}$ solve problem \eqref{eq:prob} with constraint \eqref{eq:cons}. 
 We denote the optimal value of problem \eqref{eq:prob} with constraint \eqref{eq:cons} by $V(\overline\Pi)$. Then, it follows that 
 $$\begin{aligned}
V(\overline\Pi) & =\sup _{\substack{d_1,d_2 \in [0,x_p] \\
d_2-d_1\leq  \overline \Pi}} \mathrm {MD}_g^D(T_I)
\leq \sup _{\substack{d1,d_2 \in [0,x_p] \\
d_2-d_1\leq  \overline \Pi}} \left\{\mathrm {MD}_g^D(T_I)-\lambda^*_2\left( d_2-d_1-\overline \Pi \right)\right\}  \\
& \leq \sup _{d_1,d_2 \in [0,x_p]}\left\{\mathrm {MD}_g^D(T_I)-\lambda^*_2\left( d_2-d_1-\overline \Pi \right)\right\}   =\mathrm {MD}_g^D(T_{I_{\widetilde d_{1\lambda^*_2},\widetilde d_{2\lambda^*_2}}}) \leq V(\overline\Pi).
\end{aligned}$$
 The last inequality is because $I_{\widetilde d_{1\lambda^*_2},\widetilde d_{2\lambda^*_2}}$ is feasible to problem \eqref{eq:prob}  without the constraint.  Hence, $(\widetilde d^*_1,\widetilde d^*_2)=(\widetilde d_{1\lambda^*_2},\widetilde d_{2\lambda^*_2})$  solves problem \eqref{eq:wF3}.
Thus,  we have  $\Pi((X-(\widetilde d^*_2-\widetilde d^*_1))_+)=\overline \Pi$. By letting \begin{equation}\label{eq:la1}\lambda_1=g'\left(\int _{ 0 }^{d_{1}} h_1(S_X(x))  \d x+ \int _{ d_{2}}^{x_p} h_1(S_X(x))  \d x\right),\end{equation} we can solve 
 $\widetilde d^*_1$,  $\widetilde d^*_2 $ and $\lambda^*_2$  by 
  \begin{equation}\label{eq:set1}\widetilde d^*_{1} =\inf\{x:~(1+\lambda_2-S_X(X))-\lambda_1h_1(S_X(x))\leq 0 ~~\text{and} ~~0\leq x\leq x_p\},\end{equation} and  \begin{equation}\label{eq:set2}\widetilde d^*_2 =\sup \{x:~(1+\lambda_2-S_X(X))-\lambda_1h_1(S_X(x))\leq 0 ~~\text{and} ~~\widetilde d^*_1\leq x\leq x_p\}\end{equation}  and \begin{equation}\label{eq:la2}\lambda_2= \inf \left\{ \lambda_2: \widetilde  d^*_2-\widetilde  d^*_1-\overline \Pi \leq 0, ~\text{and}~\lambda\geq0\right\}.\end{equation} 

\underline{Case 2: $\Pi=\ES_p$}. Let   $h_2(t)=\frac{t}{1-p} \wedge 1$ with $p\in(0,1)$. 
 \begin{itemize}
 \item[(i)] For  $t<1-p$, or equivalently, $x>x_p$, we always have $H(x)=\frac{p+\lambda_2}{1-p}S_X(x)- \lambda_1 h_1(S_X(x)).$ When $H(x_p)= p+\lambda_2-\lambda_1 h_1(1-p)>0$,  if $H'(M)=\frac{p+\lambda_2}{1-p}-\lambda_1h'_1(0)<0$, then there exists a unique $d_{3, \lambda_1,\lambda_2}$ such that $H_1(x)>0$ for $x_p<x<d_{3, \lambda_1,\lambda_2}$, and   $H_1(x)<0$ for $d_{3, \lambda_1,\lambda_2}<x<M$;   if $H'(0)>0$, then $H(x)\geq 0$ for any $x\in[x_p,M]$.  When   $ H(x_p)=p-\lambda h_1(1-p)<0$, then we have $H_1(x)\leq0$  for any $x>x_p$.
\item[(ii)]  For  $t\geq1-p$, or equivalently,  $x\leq x_p$, the analysis is similar to the case of VaR. 
\end{itemize}
Define   \begin{equation}\label{eq:set33}d_{3,\lambda_1,\lambda_2} =\sup\left\{x:~\frac{p+\lambda_2}{1-p}S_X(X)-\lambda_1h_1(S_X(x))\geq 0, ~~\text{and} ~~x_p<x<M\right\}\end{equation} with $\lambda_1$ given by \eqref{eq:la1}.  
 It is clear that $d_{1,\lambda_1,\lambda_2}\leq d_{2,\lambda_1,\lambda_2}\leq x_p\leq d_{3,\lambda_1,\lambda_2} .$ 
 Then problem \eqref{eq:wf3} can be  minimized by $I(x)=I_{d_{1,\lambda_1,\lambda_2},d_{2,\lambda_1,\lambda_2},d_{3,\lambda_1,\lambda_2}}(X):=\int_0^x q(t)\d t =(x-d_{1,\lambda_1,\lambda_2})_+\wedge (d_{2,\lambda_1,\lambda_2}-d_{1,\lambda_1,\lambda_2})+  (x-d_{3,\lambda_1,\lambda_2})_+$. 
 That is,  for every $s$,  there exists an $I_{d_1,d_2,d_3}(x)=(x-d_{1})_+\wedge (d_{2}-d_{1})+  (x-d_3)_+$ that does better than any $I\in\mathcal{I}$.   Next, we  aim to find the optimal $d_1$, $d_2$ and $d_3$ for problem  \eqref{eq:wf3} subject to \eqref{eq:cons} when  the insurance contract is given by $I_{d_1,d_2,d_3}$ for some $d_1, d_2\in[0,x_p]$ and $d_3\in(x_p,M]$, that is, 
 \begin{align*} \min _{d_1, d_2 \in [0,x_p], d_3\in(x_p,M]} \widetilde   F(d_1,d_2,d_3):= & \int _ {d_{1} }^{d_{2}} (1+\lambda_2-S_X(x)  \d x  +\int _ {d_{3}}^{M } \frac{p+\lambda_2}{1-p}S_X(x)   \d x \\& + g\left(\int _{ 0 }^{d_{1}} h_1(S_X(x))  \d x+ \int _{ d_{2} }^{d_{3}} h_1(S_X(x))  \d x\right)+ \E[X] -\lambda_2\overline\Pi.  \end{align*} 
By the first-order condition, we have  
\begin{align*}\frac{\partial\widetilde   F(d_1,d_2,d_3)}{\partial d_1}&=g'\left(\int _{ 0 }^{d_{1}} h_1(S_X(x))  \d x+ \int _{ d_{2} }^{d_3} h_1(S_X(x))  \d x\right) h_1(S_X(d_1))-(1+\lambda_2- S_X(d_1)),\end{align*} \begin{align*}\frac{\partial\widetilde   F(d_1,d_2,d_3)}{\partial d_2}&=-g'\left(\int _{ 0 }^{d_{1}} h_1(S_X(x))  \d x+ \int _{ d_{2} }^{d_3} h_1(S_X(x))  \d x\right) h_1(S_X(d_2))+(1+\lambda_2- S_X(d_2)),\end{align*} 
 and 
\begin{align*}\frac{\partial\widetilde   F(d_1,d_2,d_3)}{\partial d_3}&=g'\left(\int _{ 0 }^{d_{1}} h_1(S_X(x))  \d x+ \int _{ d_{2} }^{d_3} h_1(S_X(x))  \d x\right) h_1(S_X(d_3))-\frac{p+\lambda_2}{1-p}S_X(d_3).\end{align*} 
By the similar   process  of the first part,   we  can show that 
 $\Pi((X-(\widetilde d^*_2-\widetilde d^*_1))_+)=\overline \Pi$. In this case, $\widetilde d^*_1$,  $\widetilde d^*_2$,  $\widetilde d^*_3$ and $\lambda_2^*$ can be solved by \eqref{eq:set1} -- \eqref{eq:la2} and $$ \widetilde d^*_3=\sup \left\{x:~\frac{p+\lambda_2}{1-p}S_X(x)-\lambda_1h_1(S_X(x))\leq 0, ~~\text{and} ~~x_p<x<M\right\}.$$
\end{proof}


\end{document}